# Flexoelectricity and surface ferroelectricity of water ice

**Authors:** X. Wen[1,3*], Q. Ma[1], A. Mannino[4,5], M. Fernandez-Serra[4,5], S. Shen[1*], G. Catalan[2,3*]

**Affiliations:**

[1]State Key Laboratory for Strength and Vibration of Mechanical Structures, School of Aerospace Engineering, Xi'an Jiaotong University, Xi'an, China.

[2]ICREA—Institucio Catalana de Recerca i Estudis Avançats, Passeig Lluís Companys 23, Barcelona, Catalonia

[3]Institut Catala de Nanociencia i Nanotecnologia (ICN2), CSIC and The Barcelona Institute of Nanoscience and Technology (BIST), Campus Universitat Autónoma de Barcelona, Bellaterra, Catalonia.

[4]Physics and Astronomy Department, Stony Brook University, Stony Brook, New York 11794-3800, United States

[5]Institute for Advanced Computational Science, Stony Brook University, Stony Brook, New York 11794-3800, United States

*Corresponding author: xin.wen@icn2.cat; sshen@mail.xjtu.edu.cn; gustau.catalan@icn2.cat

**Abstract:**

The phase diagram of ice is complex and contains many phases, but the most common (frozen water at ambient pressure, also known as Ih ice) is a non-polar material despite individual water molecules being polar[1,2]. Consequently, ice is not piezoelectric and cannot generate electricity under pressure[3]. On the other hand, the coupling between polarization and strain gradient (flexoelectricity) is universal[4], so ice may in theory generate electricity under bending. Here we report the experimental demonstration that ice is flexoelectric, finding a coefficient of 1.14±0.13 nC/m, comparable to that of ceramics such as $SrTiO_3$, $TiO_2$, or $PbZrO_3$. Additionally, and unexpectedly, the sensitivity of flexoelectric measurements to surface boundary conditions has also revealed a ferroelectric phase transition around ~160K confined in the near-surface region of the ice slabs. The electromechanical properties of ice may find applications for low-cost transducers made in-situ in cold and remote locations. Importantly, there are also consequences for natural phenomena. In particular, we have calculated the flexoelectric charge density generated in ice-graupel collisions, and found it to be comparable to the experimental charge transferred in such events, suggesting a possible participation of ice flexoelectricity in the charging up of thunderstorms.

**Main Text:**

Ice, formed by hydrogen bonding of water molecules ($H_2O$), is one of the most widespread and abundant solids on earth. Manifesting as snowflakes, frosts, and glaciers in nature, it plays an essential role in geology, meteorology, and astronomy[1]. Moreover, even liquid water is ice-like when nano-confined or at the interface with solids[5], which renders the physics of ice also relevant to life science[6] and electrochemistry[7]. Despite the ongoing interest and large body of knowledge on ice[3], new phases[8] and anomalous properties[9] continue to be discovered, suggesting that our understanding of this ubiquitous material is incomplete.

An interesting open question concerns the electromechanical properties of ice. Despite the polarity of individual water molecules, common ice Ih is not piezoelectric[3], due to the geometric frustration introduced by the so-called Bernal-Fowler rules[1,3]: two hydrogen protons must be adjacent to each oxygen atom, but there can only be one hydrogen proton between two oxygen atoms. As a result, in contrast to the oxygen atoms, which are arranged in a hexagonal lattice, the hydrogen atoms do not exhibit long-range order[3], resulting in randomly oriented water dipoles and thus no macroscopic piezoelectricity. Yet ice is known to generate electricity under mechanical stress in nature. For example, ice collisions and fractures cause electrifications in clouds[10,11] and polar regions[12,13]. The underpinning electromechanical mechanism of these natural phenomena, however, remains elusive.

In this context, we bring our attention to flexoelectricity, a coupling between electrical polarization and strain gradients that, contrary to piezoelectricity, can exist in materials of any symmetry[4]. In theory, then, it may also exist in ice. Yet, despite a growing awareness of flexoelectricity and its consequences in other materials[14-21], the flexoelectricity of ice remains unknown. In this article we report its measurement and examine some of its repercussions.

We have prepared ice capacitors by freezing at ambient pressure a layer of ultrapure water (milli-Q, resistivity >10 MOhm·cm) between two gold-coated aluminum electrodes (Fig. 1a and b). Our samples are polycrystalline (mean grain size ~77 μm) and are at ordinary Ih phase with a preferential orientation of [001] throughout the experimental temperature range, as verified by X-ray diffraction and Raman spectroscopy (Supplementary Fig. S1). Piezoelectric measurements confirm the non-piezoelectric nature of our ice samples (Supplementary Fig. S2).

To measure ice flexoelectricity, we have used a dynamic mechanical analyzer (DMA) to deliver an oscillating three-point bending deformation to the ice capacitors (Supplementary Fig. S3). The vertical deflection (measured by a displacement sensor) and the bending-induced charge (collected by a charge amplifier) are synchronously recorded by an oscilloscope, and are then used to calculate strain gradients and polarization respectively. Fig. 1c plots the Fourier-filtered first-harmonic data of the applied displacement and the induced charge. The induced polarization, measured as a function of applied alternating strain gradient, is plotted in Fig. 1d. It shows the linear dependence expected for a flexoelectric material. The slope of the linear fit (1.65 nC/m) represents the effective flexoelectric coefficient $\mu_{13}^{eff}$ of ice Ih. The results shown in Fig. 1c and d were obtained at 237 K, a temperature well below melting, and around which ice flexoelectricity was found to be relatively temperature-independent and far from any anomaly. However, there are anomalies both at higher temperatures (>248 K) and lower ones (<203K), which we discuss next.

The temperature dependence of flexoelectricity was measured in the range between 143 K and 273 K simultaneously with the mechanical response of the ice slabs. Representative results for three samples are plotted in Fig. 2a. Three distinct regimes can be identified. Above 248 K, flexoelectricity starts to increase, and the phase angle between strain gradient and induced polarization shifts away from 0 deg to 180 deg (the sign of $\mu_{13}^{eff}$ turns negative, Fig. 2a and b), accompanied by the onset of mechanical creep (Fig. 2c). These anomalies coincide with the surface transition to pre-melting quasi-liquid layers (QLL)[11,22] (see Supplementary Fig. S4 for direct comparison between reported QLL results and our measurements). The QLL of ice contains mobile ions that facilitate charge transport[11] and grain-boundary sliding[23], consistent respectively with the observed increase of flexoelectricity[16] (and shift its phase angle from capacitive to conductivity-dominated response[24]), and with the mechanical softening above 248 K (Fig. 2). The results therefore indicate that the flexoelectric and mechanical anomalies above 248 K can be attributed to the onset of pre-melting. To ensure a creep-free initial condition, measurements were always started from about 223 K (i.e. below the QLL temperature) rather than 273 K, then cooled down to 143 K (Supplementary Fig. S5), finally followed by heating measurements all the way up to 273 K (Fig. 2 and Fig. 3).

In the temperature range between 203 K to 248 K, the flexoelectric coefficient was found to be relatively constant. Taking the weighted average of all data points yields 1.14±0.13 nC/m. This is in the same range as dielectric ceramics such as $SrTiO_3$[25], $TiO_2$[16], or $PbZrO_3$[26]. Since flexoelectricity is proportional to dielectric permittivity[4], this result is consistent with the relatively high dielectric constant of bulk ice, $\varepsilon_r \sim 100$[3]. The flexocoupling coefficient (flexoelectric constant divided by dielectric permittivity) is 1.29±0.15 volts, which also agrees with the 1-10V expected range for intrinsic flexoelectricity in solids[4]. We therefore conclude that 1.14±0.13 nC/m is likely to be the intrinsic value of the effective transverse flexoelectric coefficient of ice. Note that this value is a combination of flexoelectric tensor components (i.e., $\mu_{13}$ and $\mu_{11}$ for isotropic solids)[4].

Below 203 K, however, the flexoelectric coefficient begins to grow again, reaching a peak of ~7.6 nC/m at around 164.6 ± 1.7 K (Fig. 2a). Such a temperature dependence of flexoelectricity has only been observed before in ceramic materials with ferroelectric or antiferroelectric phase transitions[26-28], so a question emerges: is the flexoelectric anomaly of ice related to a ferroelectric phase transition?

Ice can become ferroelectric in some conditions. Doping bulk ice with alkali hydroxides causes proton rearrangement and transform ice-Ih to ferroelectric ice-XI at 72 K[1]. On the other hand, the temperature of the flexoelectric anomaly of our ice slabs is ~93 K higher than the Curie temperature of bulk doped ice. Moreover, the mechanical response of our samples (Fig. 2c) shows no sign of any structural phase transition, nor do the Raman measurements as a function of temperature of our samples show any change of symmetry (Supplementary Fig. S1), and the maximum bending-induced stress in the surfaces of our sample is ~0.005Gpa (Supplementary Fig. S6), two orders of magnitude smaller than the minimum stress that can cause a phase transition in ice [29,30]. The flexoelectric peak at 164.6 K therefore cannot be attributed to a bulk or strain-induced ferroelectric transition.

On the other hand, ultra-thin films of ice grown on platinum substrates have been reported to become ferroelectric[31,32] around 163 K~175 K[32,33], which is similar to the temperature range of our flexoelectric anomaly. Although our ice samples are thick (~2 mm) and have gold electrodes instead of platinum, the similarity of critical temperatures is tantalizing. A unique feature of flexoelectricity is in fact that it contains comparable contributions from the bulk and the surface, irrespective of the sample thickness[34,35]; even in bulk samples, surfaces can still determine the magnitude and even the sign of the total flexoelectric coefficient[16,24,36,37]. Given the inherent contribution of surfaces to total flexoelectricity, the lack of evidence for a bulk phase transition, and the known existence of ferroelectricity in thin films with a similar critical temperature as that of our flexoelectric peak, we hypothesize that the origin of the flexoelectric peak is a ferroelectric phase transition confined within the near-surface region (aka "skin layer") of our samples. Skin layers with distinct properties are common in oxide electroceramics[38,39], and ice itself is already known to show an interfacial QLL transition below the bulk melting temperature[22], so it is not unreasonable to expect the surface of ice to have its own ferroelectric transition with a $T_c$ different from the bulk. Next, we examine this hypothesis in closer detail.

In a slab with opposite polar surfaces, surface piezoelectricity can contribute to effective flexoelectricity as depicted in Fig. 3a: under bending, the different signs of the strain (compressive on one side, tensile on the other) cause a difference in polarization, which is additional to the bulk flexoelectricity. Within this model, and assuming the piezoelectric coefficient of ice XI for the interfacial layers, we expect their interfacial layer thickness ($t_i$) to be $t_i(H_2O/Au)$=20.3 nm (Supplementary S3, 4). It has been suggested[40] that the free surface of ice remains polar at all temperatures below the premelting point (~248K in our sample), but such a result cannot explain the observed phase transition at 164.6 K. The difference is likely related to the presence of electrodes in the ice capacitors. To test this hypothesis, we have replaced the Au electrodes with other metals. Measurements on three different Pt/ice/Pt capacitors are presented in Fig. 3b, showing similar temperature dependence of flexoelectricity but a larger maximum of ~15 nC/m at ~ 158.9±2.6 K (see Supplementary Fig. S8 for phase angle). The flexoelectric peak of ice with Pt electrodes is larger than that with Au electrodes (Fig. 3b). The calculated thickness of the piezoelectric interfacial layers in this case is $t_i(H_2O/Pt)$=31.3 nm (Supplementary S4).

The electrode effect is consistent with Sugimoto's proposal for the origin of ferroelectricity in thin films[32], whereby interfacial ferroelectric ordering is triggered by electron transfer from the surface to the electrode in order to equilibrate chemical potentials. The work function $\emptyset$ of gold (~5.1 eV)[41] is higher than that of ice (~4.4 eV)[42,43], and the work function of Pt (~5.65 eV)[41] is higher yet, resulting in a bigger poling field, consistent with the observed higher flexoelectric anomaly. Conversely, the work-function of $Al_2O_3$ (the oxidized layer on Al electrode, Supplementary Fig. S9) is about 4.26 eV[44], smaller than those of Au and Pt and very close to that of ice. The interfacial field between ice and Al electrode should be very small, and indeed the ice capacitors with Al electrodes show no visible flexoelectric peak (Fig. 3b). In Fig. 3c we plot the peak value of the flexoelectric coefficient as a function of the electrode work-functions, showing a direct proportionality.

We have further studied the effect of the pre-poling field on the flexoelectric coefficient and found a butterfly hysteresis loop (Fig. 3d). Note that each point of the loop was measured after switching off the biasing voltage, meaning that the hysteresis is not due to a leakage artifact[45]. The measurement of hysteresis, which is a defining feature of ferroelectricity, had been elusive in ice prior to the present work, due to the difficulty of depositing non-shorting electrodes in ultra-thin films such as Sugimoto's. Besides supporting the existence of interfacial ferroelectricity, therefore, the present results also showcase how flexoelectricity can be utilized to explore interfacial properties while retaining the convenience of bulk samples.

To further understand the microscopic origin of the surface phase transition, we also computed the surface free energies of ice-metal interfaces, using density functional theory. It has been previously shown, both experimentally and theoretically[31,46-48], that metallic Au, Pt, and Pd [111] surfaces affect the dipole orientation of $H_2O$ at the interface. Although our samples are polycrystalline, X-ray diffraction (Supplementary Fig. S1) shows that the predominant out-of-plane orientation of the ice grains is hexagonal [001], and cubic [111] for the electrodes, so we assume these orientations in the calculations. We have computed the free energies of two interfaces: ferroelectric ice-XI on Au[111], with the ice-XI dipole pointing towards the metal surface (Fig. 4a), and common ice-Ih on Au[111], without ordered dipoles (Fig. 4b). We then compare the difference in cohesive energy between these two systems with the cohesive energy difference between bulk ice XI and bulk ice Ih without the Au slab. The result is that the Au interface enhances the stability of the polar phase by 140 meV.

When the interfacial energy gain is distributed among all the molecules in the interfacial layer and subtracted from the Helmholtz free energy of ice XI (Fig. 4c), the calculations predict that the Curie temperature should shift to the observed phase transition temperature of 164.6 K for a skin layer thickness of 14.7 nm (Fig 4d, Supplementary S5), comparable to our earlier experimental estimate of $t_i(H_2O/Au)$=20.3 nm. We have also repeated the calculations replacing Au with Pt, finding that the stability of the proton ordered (XI) interface relative to the proton disordered (Ih) interface is enhanced by 307 meV. At the observed phase transition temperature of 158.9 K (for the ice capacitors with Pt electrodes), this corresponds to a ferroelectric skin layer thickness of 34.7 nm (Fig. 4d), in good agreement with the experimental estimate of $t_i(H_2O/Pt)$=31.3 nm. Further details of the calculations are provided in the supplementary materials.

Lastly, we examine some consequences of ice flexoelectricity in nature. It has long been known that collisions between ascending small ice particles and descending large graupel particles cause charge separation in clouds (Fig. 5a), and eventually cause lightning[10,11]. Despite numerous existing theories[10,11] (see Supplementary S7), the mechanism behind such charge separation remains elusive due to its inherent complexity. In this context, flexoelectricity has been recently suggested to be at play in triboelectricity in other material systems[49-55]. Now that we have detected and measured the flexoelectricity of ice, and considering that strain gradients are always generated during collisions, we examine whether collision-induced flexoelectricity, without additional parameters, is commensurate with the known contact electrification of ice.

We have calculated the induced flexoelectric polarization in a typical ice-graupel collision (see details in Supplementary S8~10). Fig. 5b shows the distribution of polarization corresponding to the maximal deformation during impact, which reaches ~$10^{-4}$ C/m$^2$ on the graupel surface while remains much smaller on ice surface, primarily because graupel is softer and more deformable than pure ice. This surface polarization creates a depolarizing field (~$10^5$ V/m, Fig. S13a) that will attract positive free charge. This may be provided by ions in the surface QLL of ice (see also Supplementary S6)[11,56]. Since graupel brings more positive charges towards the contact interface than ice does, their physical disengagement presumably leaves graupel negatively charged and ice positively charged, as illustrated in Fig. 5a.

Note that this scenario corresponds to a positive flexoelectric coefficient. When the coefficient turns negative (as observed upon temperature increase in Fig .2b), the predicted direction of charge separation would reverse, leaving graupel positively charged and ice negatively charged. This agrees with prior experimental evidence of temperature-driven polarity reversal[10,57,58], a phenomenon considered as the origin of thunderstorm's tripole structure[59]. Besides determining the sign of charge transfer, we can go further in the quantitative evaluation of its magnitude. At the moment of maximum indentation, assuming that all the flexoelectric polarization is screened by free charges, the integrated flexoelectric polarization over the contact area represents the upper bound of interfacial free charge $Q$ that may be directly attributed to flexoelectricity:

$$Q \approx 15.486\mu_{13}^{\mathrm{eff}} R_i v_r^{\frac{4}{5}} \rho^{\frac{2}{5}} \frac{Y_i - Y_g}{(Y_i + Y_g)^{\frac{3}{5}} (Y_i Y_g)^{\frac{2}{5}}} \quad (1)$$

where $R_i$ is the radius of ice particles, $v_r$ is the impact velocity, $\rho$ is the density of ice, $Y_i$ and $Y_g$ are Young´s modulus of ice and graupel particles respectively (see derivations in Supplementary S9). Plotting $Q$ as a function of $R_i v_r{}^{4/5}$, we see that the theoretical prediction matches experimentally reported values of contact charge transferred per ice-graupel collison[58,60-66] (see Supplementary S10 for the details of each experimental dataset used).

Despite these promising agreements, we make no claim to have fully solved the long-standing ice-charging problem. Our model is highly simplified and, as mentioned, it can provide only an upper bound, i.e. it calculates how much flexoelectrically-generated free charge upon contact, but it does not calculate how these free charges will be distributed upon release. Note, in particular, that the model requires that the free charges do not return to their origin during release, i.e. they must become trapped, either by defects, or by a metastable (anelastic, plastic or viscous) status of the collision-induced deformation. These additional complexities — plasticity, fracture, hydrodynamics, impurities, mass transfer, and contact-associated phase transitions— are all factors that are difficult to quantify at this point. Meanwhile, flexoelectricity does not exclude other charging mechanisms and may even interact with them in subtle ways (Supplementary S6). Detailed caveats and suggestions are provided in Supplementary S12. We encourage future systematic research to complete the present model and fully elucidate the role of flexoelectricity in thunderstorm electrifications.

In conclusion, water ice is flexoelectric and, at its "skin layer", ferroelectric below ~160K. Most importantly, flexoelectricity makes ice electromechanically active at all temperatures, and will therefore participate in any natural process involving mechanical deformations of ice or ice-like interfacial water.

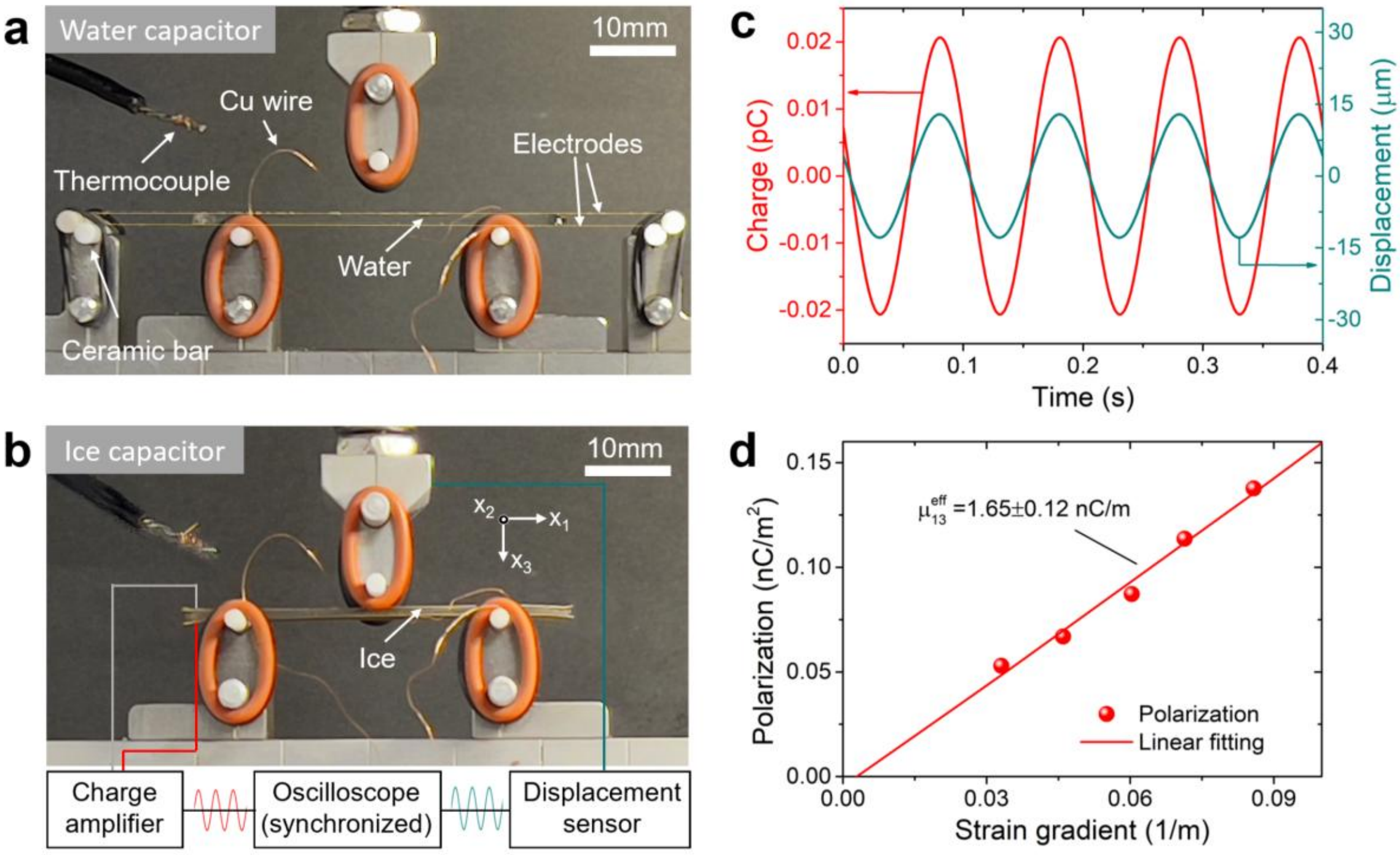

**Fig. 1 Experimental setup for measuring ice flexoelectricity.** **a**, A water capacitor, consisting of a layer of water and two pieces of Au electrodes (highlighted by yellow lines). **b**, An ice capacitor placed in the DMA for oscillating three-point bending deformation. The displacement in the center and the bending-induced electric charge are recorded by an oscilloscope synchronously. **c**, Fourier-filtered first-harmonic displacement (the cyan curve) and charge (the red curve) for an applied displacement of 12.5 $\mu$m. **d**, Electric polarization versus strain gradients, and linear fit to the results. The data in (**c**) and (**d**) was obtained at 237 K.

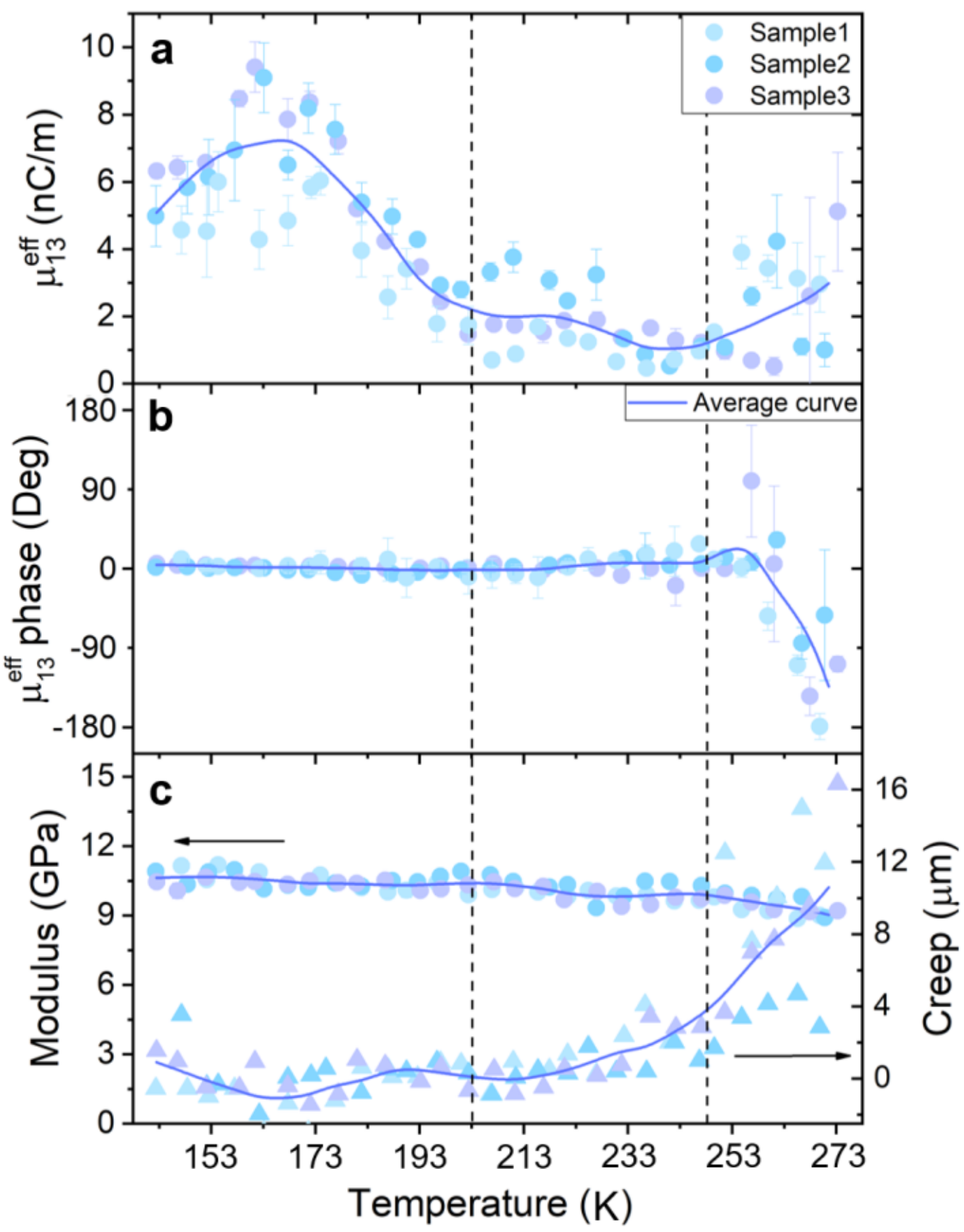

**Fig. 2. Temperature dependence of ice flexoelectricity and mechanical properties. a**, The effective flexoelectric coefficient as a function of temperature. **b**, The phase angle between displacement and polarization charge as a function of temperature. **c**, The modulus and the creep displacement in the first ten seconds of loading as a function of temperature. The data shown in this figure is obtained in three samples with Au electrodes. The solid lines represent the smoothed average curve. The error bars in (a) and (b) represents the standard error from linear regressions and averages respectively.

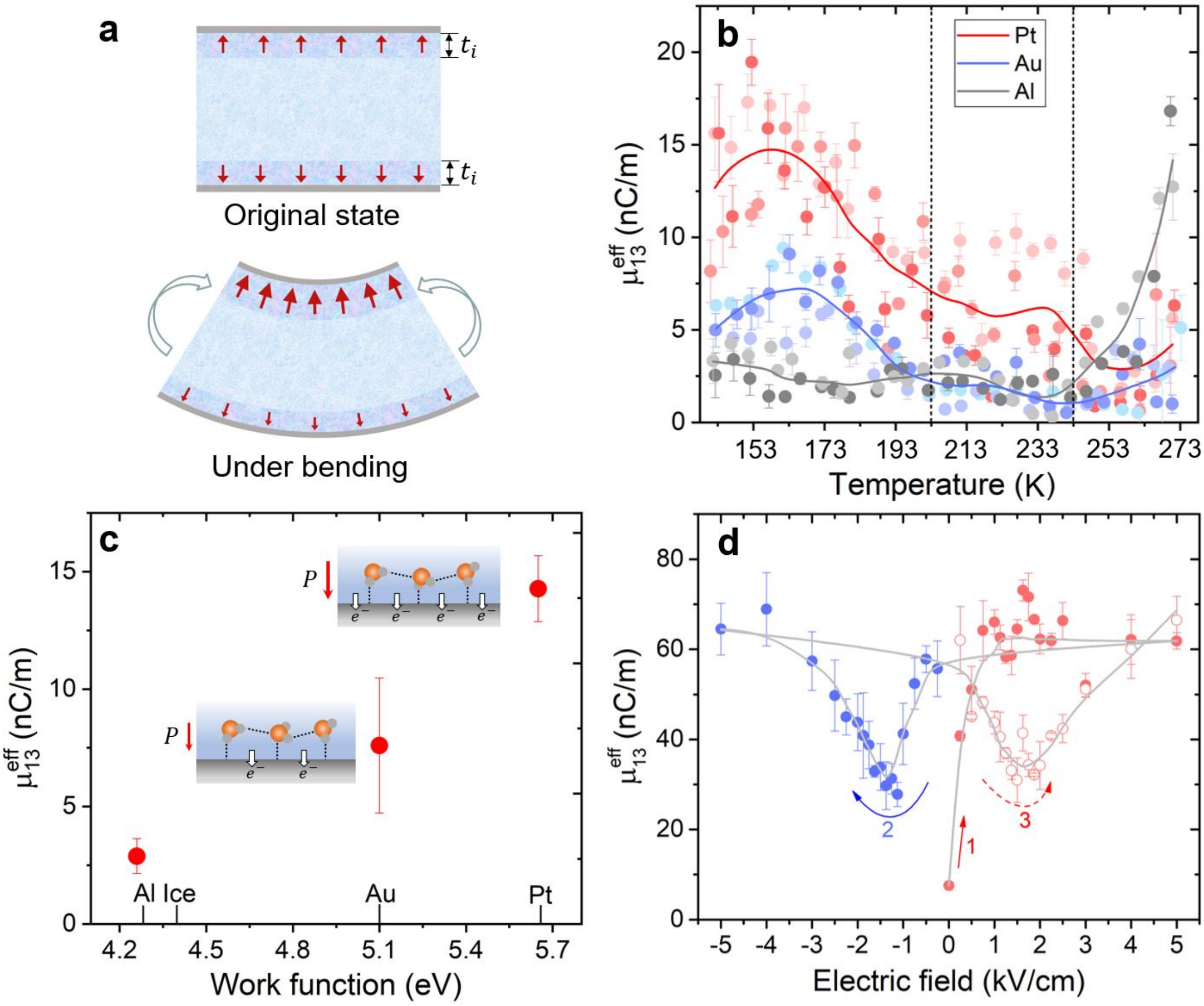

**Fig. 3. Surface contribution to enhanced flexoelectricity. a,** Schematic illustration of how surface polarization can contribute to effective flexoelectricity in a slab with polar surface layers. **b,** Experimental flexoelectric coefficient of ice with Pt, Au, and Al electrodes as a function of temperature. The solid lines are a smooth average of all the capacitors having the same type of electrode. The electrode-dependent results indicate a dependence of flexoelectricity on surface boundary condition. **c,** Value of the flexoelectricity at 164.6 K as a function of the electrode work functions. Note that Al electrodes have an oxidized layer of $Al_2O_3$. **d,** The evolution of flexoelectricity as a function of poling electric field measured at ~155K and with Au electrodes. The error bars in (b) represents the standard error from linear regressions. The error bars in (c) and (d) represent the standard error from averages of (c) different samples or (d) multiple measurements.

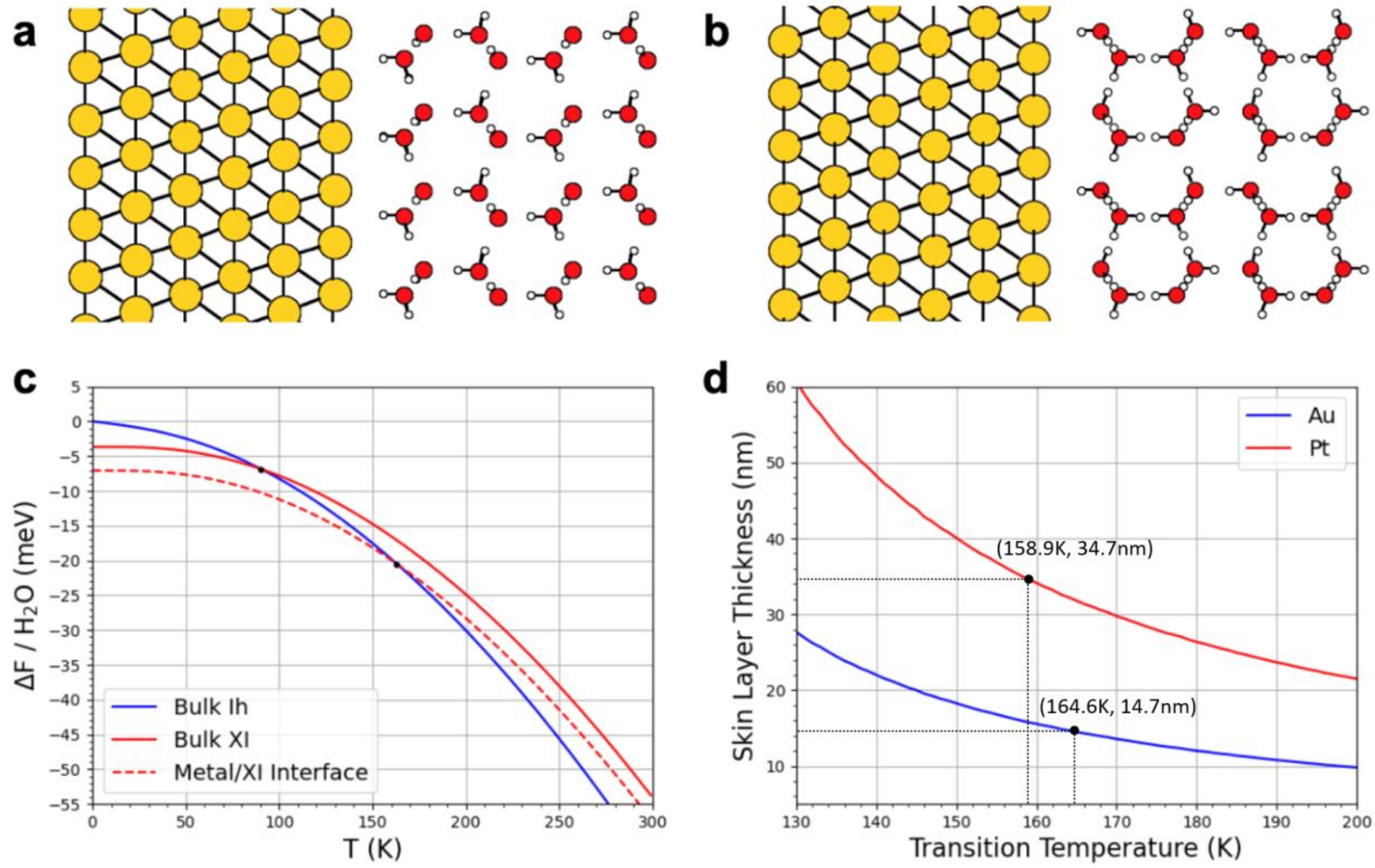

**Fig. 4.** ***Ab initio*** **simulations of metal [111]-ice (Ih and XI) [001] interfaces. a,** Illustration of the Ice XI / Au[111] system. **b**, Illustration of the Ice Ih / Au[111] system. **c,** Relative free energy for ice XI and ice Ih, showing that the transition temperature $T_c$ is increased within the metal/ice interface than that in the bulk ice[67]. **d,** Relationship between the increased $T_c$ and the skin layer thickness for Au[111] and Pt[111].

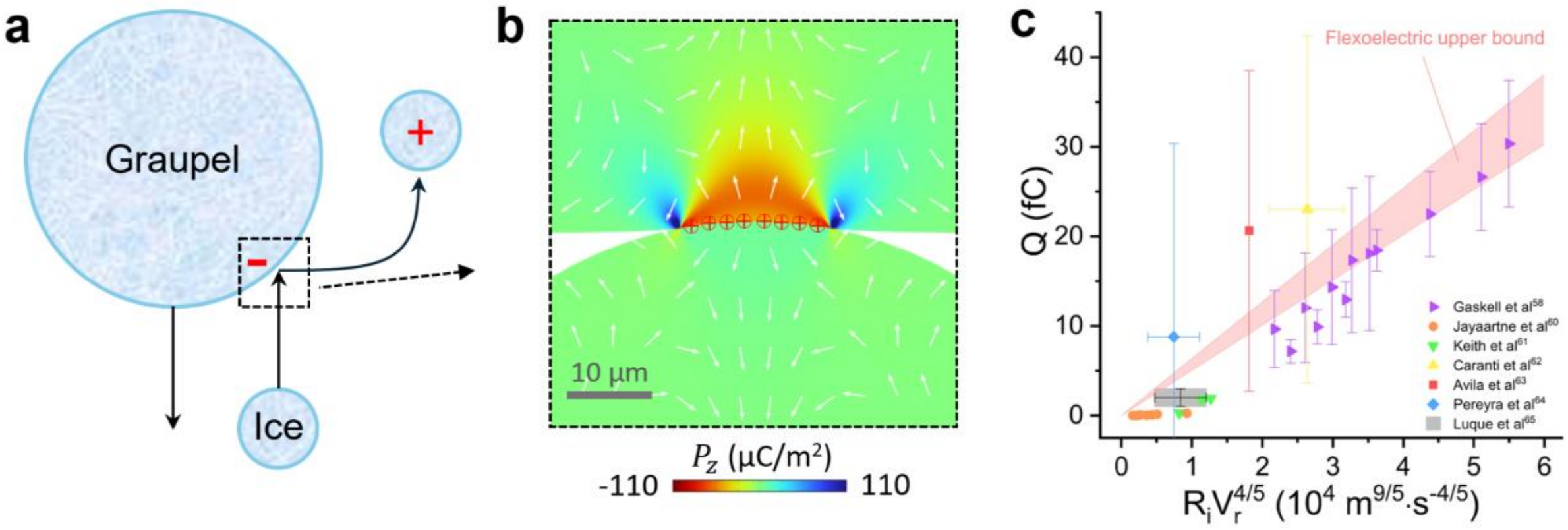

**Fig. 5. Flexoelectricity in ice electrification events. a,** Schematic illustration of charge separation in a typical ice-graupel collision in clouds (adapted from[3,10,68]). **b,** Calculated flexoelectric polarization $P_z$ near the contact interface between the graupel and ice particle, corresponding to panel a. Arrows of normalized length indicate the direction of the polarization vector. Positive free charges, presumably from the quasi-liquid layers on surfaces, are attracted to screening the polarization. **c**, The dependence of transferred charge on $R_i v_r^{\frac{4}{5}}$ per ice-graupel collision, showing both the predicted flexoelectric upper bound and the experimental measured values in previous studies[58,60-66]. The pink band denotes the uncertainty range of the adopted flexoelectric coefficient (1.14±0.13 nC/m). See Supplementary Table S1 for parameter details.

## Methods

**Preparation of electrodes.** Aluminum foils with a thickness of 15 $\mu$m (BS-QT-027, Biosharp) were cut into strips with a length of 100 mm and width of 5 mm, which were then chemically cleaned by acetone, ethanol, and ultrapure water sequentially. To study the electrode dependence of ice flexoelectricity, we coated a gold or platinum layer of ~100 nm on the surface of cleaned Al foils with an Ion Sputter Coater (MC1000, HITACHI). Copper wires were attached to the foils by a drop of silver paste (948-06G, HumiSeal), which was then solidified at 373 K for half-hour. To check the conductivity of the adhesion, we used a multimeter to measure the resistance from the copper wire to the aluminum foil. When the resistance is less than 1 Ohm, the electrodes were used for the following steps.

**Preparation of water and ice capacitors.** Ultrapure water produced by a water purification system (Sistema Milli-Q Advantage A10) was used as the mother material for fabricating pure ice. Water was degassed in a vacuum drying oven (DZF-6050, ShanghaiYiheng) at room temperature until no air bubbles were generated. Two pieces of cleaned aluminum foil (or sputtered with Au or Pt) were placed horizontally on the dynamic mechanical analyzer with a vertical air gap. The air gap was engineered by placing two ceramic bars on the left and right sides of the clamps as shown in Fig. 1a. Degassed ultrapure water was then added into the air gap between foils by a plastic dropper. Thanks to the surface tension, the water could maintain a good shape instead of spreading to the left and right sides. With two electrodes and one layer of water in between, a water capacitor was obtained (Fig. 1a). Then, the water capacitor was frozen at 253 K for an hour and a half to obtain the ice capacitor. By cutting the redundant electrodes on left and right sides, we had an ice capacitor ready for flexoelectric characterizations (Fig. 1b). The thickness of the ice capacitors was measured by a vernier caliper in the temperature box after the flexoelectric measurements, which ranges from 1.8mm to 2.2mm.

**Structural characterizations.** The ice surface (Fig. S1a) was imaged under an optical microscope (NIKON ECLIPSE LV100D) with a temperature chamber (Linkam, HFS600E-PB4). Raman spectrum (Fig. S1c) was measured by a confocal Raman spectrometer (Witec Alpha300R) with the same Linkam chamber. The laser wavelength, power density, and grating were 532 nm, 8 mW and 600 lines/mm$^{-1}$ respectively. X-ray diffraction patterns (Fig. S1d) were measured by a Diffractometer (X'pert Pro MPD Malvern-Panalytical) with a low-temperature chamber from Anton Paar (TTK600). The top electrode of the ice capacitor, grown in-situ in the temperature chamber using the same method described above, was removed from ice to enable these measurements.

**Flexoelectric characterizations.** A dynamic mechanical analyzer (Electroforce 3200, TA Instruments) was used to apply an oscillating 3-point bending deformation, which was recorded by a High Accuracy Displacement Sensor (HADS) embedded in the DMA. To reduce 1/f noise from the environment, relatively high drive frequencies (10~17 Hz) of the dynamic force were used, which is still far from the resonant frequency and so the loading can be regarded as a quasi-static condition. A small static force (typically -1.25N) was applied simultaneously to hold the sample in place.

The AC displacement $\delta$ at the sample's center measured by the HADS is converted to the average strain gradient across the electrode area by[1]

$$\overline{\frac{\partial \varepsilon_{11}}{\partial x_3}} = \frac{12\delta}{L^3}(L-a), \tag{1}$$

where $a$ is the half-length of the electrodes, $L$ is the distance between two bottom ceramic bars. In our experiments, $L$ and $a$ are 30mm and 15mm respectively. The bending-induced electric charge was converted by an amplifier (2692, Brüel & Kjær) to a voltage signal by a charge gain of 100mV/pC, which was then recorded with the bending displacement synchronously by an oscilloscope (MDO3104, Tektronix). The measured charge $Q$ was converted to the average electric polarization across the electrode area along the thickness direction by

$$\overline{P_3} = \frac{Q}{A}, \tag{2}$$

where $A$ is the area of the electrodes and equal to 30mm*5mm in our experiments. To exclude the environmental interferences, we performed the Fourier transformation on the recorded displacement $\delta$ and charge $Q$ and used the first-harmonic signal for the calculations in equation (1) and (2). With the measured strain gradient and polarization, the effective flexoelectric coefficient is calculated by

$$\mu_{13}^{eff} = \overline{P_3} / \overline{\frac{\partial \varepsilon_{11}}{\partial x_3}} \tag{3}$$

To further make sure what we are measuring is flexoelectricity rather than the environmental noise, we applied different forces to the sample and measured the corresponding induced charge. By doing a linear regression of the relationship between strain gradients and polarization, we extracted the slope value as the effective flexoelectric coefficient and the standard error value as the error bar (Fig. 1d). Every data point in Fig. 2a, Fig. 3b and Fig. S4, including its mean value and error bar, is calculated by such a linear regression.

Temperature-dependent measurements were achieved by competitive action of a resistive heater and a liquid $N_2$ bath. We started the flexoelectric measurements at ~223 K, well below the temperature at which ice pre-melting and creeping plastic deformation can occur, but still well above the ferroelectric transition of the surface layer. From that temperature, we measured on cooling down to 143 K, followed by heating measurements all the way up to 273 K. To avoid the influence of the pyroelectric effect or any other interference, each measurement was performed at a fixed temperature instead of sweeping the temperature with a constant strain gradient. The temperature interval between each flexoelectric measurement was set to ~5 K and the flexoelectric coefficient at each temperature was measured by doing the least squares regression of the polarization as a function of strain gradient for five different applied strain gradients.

**Ab Initio Simulations.** We used the siesta code[2,3] to perform DFT calculations within the generalized gradient approximation (GGA) to the exchange and correlation (XC) functional. The calculations use a combination of PBE[4] and vdW-DF(PBEx)[5,6] functionals. These density functionals have previously been shown to give good results for the bulk XI-Ih phase transition[7,8]. Relaxations are done using the same combination of parameters as described in[8]. For the free energy calculations which include the nuclear quantum effects[7], the vibrational modes are calculated using the frozen phonon approximation. We only compute the phonons for the bulk ices, and add to the free energy the classical energy of the ice-Au[111] interface. The proton order-to-disorder phase transition temperature is determined from the Helmholtz Free energy at zero pressure. We account for nuclear quantum effects in the bulk ices within the quasi-harmonic approximation, with details provided in Supplementary S5(B). The main modification to the free energies of the bulk ices in this work arises from accounting for the zero-temperature interfacial binding energy contribution to the cohesive energy of ice. The computation of this energy is described in Supplementary S5(A).

**Acknowledgments:** We thank J. Liu for her code for processing the experimental data, N. Domingo, D. Pesquera, M. Stengel for useful discussions, and P. Vales, J. M. Caicedo, D. Pesquera, S. Ganguly, J. Padilla for technical support. We thank the anonymous referee #5 for suggesting and actually deriving the closed-form expression for contact charge $Q$, as well as other constructive comments. This project is funded by the Spanish Ministry of Economy, Industry and Competitiveness (projects MAT2016-77100-C2-1-P), the Catalan AGAUR agency (project: 2017-SGR-579), and the National Natural Science Foundation of China (Grant No. 12090030). ICN2 is funded by the CERCA programme/Generalitat de Catalunya and by the Severo Ochoa programme (SEV-2017-0706). X.W. acknowledges the support from the China Scholarship Council and Juan de la Cierva fellowship. M.F.-S. and A.M. were funded by the U.S. Department of Energy, Office of Science, Basic Energy Sciences, under Award No. DE-SC0019394, as part of the CCS Program.

**Author contributions:** G.C. conceived the idea and coordinated this work. G.C. and X.W. designed the experiments. X.W. and Q.M. performed the experiments under the supervision of G.C. and S.S.. A.M. and M.F-S. performed the *ab initio* calculations and associated data modeling. X.W. performed the calculations and analysis of the electrification section. X.W. and G.C. wrote the manuscript with the input from all the other authors. All authors discussed the results and commented on the manuscript.

**Competing interests:** Authors declare that they have no competing interests.

**Data and materials availability:** All data are available in the main text or the supplementary materials

Supplementary Information for

# Flexoelectricity and surface ferroelectricity of water ice

**Authors:** X. Wen[1,3*], Q. Ma[1], A. Mannino[4,5], M. Fernandez-Serra[4,5], S. Shen[1*], G. Catalan[2,3*]

**Affiliations:**

[1]State Key Laboratory for Strength and Vibration of Mechanical Structures, School of Aerospace Engineering, Xi'an Jiaotong University, Xi'an, China.

[2]ICREA—Institucio Catalana de Recerca i Estudis Avançats, Passeig Lluís Companys 23, Barcelona, Catalonia

[3]Institut Catala de Nanociencia i Nanotecnologia (ICN2), CSIC and The Barcelona Institute of Nanoscience and Technology (BIST), Campus Universitat Autónoma de Barcelona, Bellaterra, Catalonia.

[4]Physics and Astronomy Department, Stony Brook University, Stony Brook, New York 11794-3800, United States

[5]Institute for Advanced Computational Science, Stony Brook University, Stony Brook, New York 11794-3800, United States

*Corresponding author. Email: xin.wen@icn2.cat; sshen@mail.xjtu.edu.cn; gustau.catalan@icn2.cat

## S1. Excluding bending-induced phase transitions

Pressure-induced phase transition in ice is a well-known effect, but it can be excluded in our experiments, because the stress at the surface is too small compared to the energies involved in the structural phase transitions of ice. To illustrate this, we have performed a finite element simulation (COMSOL Multiphysics 5.3) using the same geometric parameters, material properties, and mechanical boundary conditions of our experiments. Fig. S6 shows the distribution of the in-plan stress $\sigma_{11}$ across the beam under the maximum force (2.5N) used in our experiments. As expected, the beam experiences compression on one side and stretching on the other side; and the maximum bending-induced stress at the surface is ~0.005GPa. This is orders-of-magnitude smaller that the pressure needed to induce a phase transition to ice IX (~0.2 GPa[1,2]) or amorphization (~1 Gpa[3]). Regarding pressure melting, ~0.005GPa only reduces the melting point by less than 0.4 K[4].

## S2. Analysis of the contribution from grain boundaries

While our samples are polycrystalline and grain boundaries are non-centrosymmetric, the grain-boundary contribution to the total flexoelectric effect can be neglected, both on theoretical and on experimental grounds. Theoretically, we must remember that at each grain boundary there are two opposing surfaces, corresponding to the adjacent grains. These surfaces have opposite orientation and, therefore, on average their surface-piezoelectric contributions will cancel each other. As indirect evidence, the flexoelectricity of $BaTiO_3$ and $SrTiO_3$ has been studied in ceramics[5,6] and single crystals[7,8], finding no significant difference of flexoelectric coefficient, suggesting that the existence grain boundaries in ceramics does not significantly affect the flexoelectric outcome. To provide direct experimental evidence for lack of grain boundary contributions, we have performed new flexoelectric experiments with different grain sizes, achieved by annealing the sample for different durations at 267 K, during which ice grains recrystallized and changed grain sizes (Fig. S7a~c). All flexoelectric measurements at 233K fell into comparable range (Fig. S7d), and no conclusive effect of grain size on ice flexoelectricity was observed.

## S3. Ab initio calculation of the piezoelectric constant of ice XI

The unit cell for this calculation is an ice XI unit cell with 4 water molecules. The exchange-correlation functional is VDW/DRSLL. The lattice vectors have the following form: $[0, a, 0]$, $\left[a\sqrt{3}/2, a/2, 0\right]$, $[0, 0, c]$ with lattice parameters a and c. First, we adjust c with a fixed lattice constant, a, and run a SIESTA calculation that allows the "out-of-plane" H atoms to relax.

We then determine the most energetically favorable configuration and repeat this for several values of a (Fig. S10a). Mulliken charges for each atom in the system are computed by SIESTA[9,10], and we use them to calculate the polarization in the c-direction. Finally, we plot the polarization as a function of the strain on the lattice constant (Fig. S10b). The slope of this plot is the transverse piezoelectric constant $e_{13}$, which was found to be 0.32 C/m$^2$.

**S4. Surface piezoelectricity**

Fig. 3a schematically shows how surface ferroelectricity can contribute to effective flexoelectricity. In a dielectric sample with polar skin layers at the surface, the surface-piezoelectric contribution to the total effective flexoelectricity is[11,12]

$$\mu_{13}^{eff} = e_{13}^{surf} \lambda \frac{h\varepsilon_b}{2\lambda\varepsilon_b + h\varepsilon_\lambda} \quad \text{(S1)}$$

where $e_{13}^{surf}$ is the transverse surface piezoelectric constant, $\lambda$ and $h$ are the thickness of the skin layer and the bulk, $\varepsilon_\lambda$ and $\varepsilon_b$ are the dielectric constant of the skin layer and the bulk respectively. When $\lambda$ is much thinner than $h$ ($\frac{\lambda}{h} \ll 1$), equation (S1) can be simplified as

$$\mu_{13}^{eff} = e_{13}^{surf} \lambda \frac{\varepsilon_b}{\varepsilon_\lambda} \quad \text{(S2)}$$

By subtracting the bulk contribution from the measured flexoelectric peak (~7.6 nC/m), we can obtain a surface-contributed $\mu_{13}^{eff}$ of ~6.5 nC/m in Au-electrodes samples. $e_{13}^{surf}$=0.32 C/m$^2$ is obtained in Section 3 by ab initio calculations. We assume that the dielectric constant of the skin layer is similar to that of the bulk, i.e., $\varepsilon_b/\varepsilon_\lambda$=1. Substituting these parameters in equation (S2), we estimate the skin-layer thickness $\lambda$ at Au/ice interface to be ~20.3 nm. The surface-contributed $\mu_{13}^{eff}$ in Pt-electrodes samples is ~10 nC/m, leading to a skin layer thickness $\lambda$ at Pt/ice interface of ~31.3 nm.

According to equation (S2), the effective flexoelectric coefficient only depends on the thickness of the skin layer $\lambda$, and is independent from the bulk sample thickness $h$. As demonstrated in previous works[13,14], the flexoelectric coefficient can only be proportional to sample thickness when the samples are semiconductors with Schottky junctions, which is not the case for ice. The independence on bulk thickness may seem counter-intuitive for a claimed surface property, but it is in fact reasonable. When a beam is bent with a given curvature, increasing sample thickness linearly reduces the surface/volume ratio but, at the same time, linearly increases the surface strain. The two opposite dependencies mutually cancel, resulting in no bulk thickness dependence. This conclusion has been affirmed in the theoretical literature[11,12] and is backed by our own flexoelectric measurements for samples with different thicknesses (from 1mm to 2mm), as shown in Fig. S11 below.

**S5. Ab initio calculation of relative energy between ice-metal interface**

A. Cohesive energy calculations at T=0.

Simulating these systems and being able to compare them is not trivial due to the polar nature of the ice XI phase. The limitations on the size of the system that can be computationally accessible poses a general problem when studying water or ice interfacial properties using ab initio simulations. In the case of ice XI, two different options to study the interface are presented in Fig. S12. These are non-ideal.

(a) two different interfaces with the metal are computed in one calculation. One with the protons facing the metal (positive end of the polarized slab), and one with the O atoms facing the metal (negative end of the slab). The polarization is screened by the metal, but it is difficult to separate the contributions from the two interfaces. Alternatively, it is possible to create a proton ordered slab with symmetric interfaces as in

(b) both interfaces with the presumed favorable H-Metal interface. To achieve this, one needs to create a layer of H-bond defects in the middle of the slab. These water molecules do not completely satisfy Bernal–Fowler–Pauling ice rules (There is always a proton in between each pair of oxygen atoms and each oxygen always keeps two close, covalently bound protons).

When comparing these structures to a proton disordered system (which always has, by definition, symmetric interfaces), the proton disordered system will always be more stable. Hence it is necessary to account for the defect formation energy, and this makes the calculation much more complicated. We can calculate the cohesive energy per water molecule with the following equation:

$$E_{\mathrm{Cohesive/H_2O}} = E_{\mathrm{Ice/Metal}} - E_{\mathrm{(Ice-H_2O)/Metal}} - E_{\mathrm{H_2O}} \tag{S3}$$

with the subscripts denoting which species are included in the calculation. In the second term, we remove a single water molecule from the surface layer of the ice. The unit cell and all other simulation parameters remain the same.

This calculation is done for system (b) and a similar system (c) with ice Ih instead of ice XI. The difference in cohesive energy for these two systems is 233 meV, with system (b) being more favorable. We also do this calculation for bulk ice XI and bulk ice Ih without the Au slab. The difference in cohesive energy for these two systems is 93 meV, with the bulk ice XI system being more favorable. We conclude from this that the Au interface creates an enhanced stability ($\Delta E_{\mathrm{Surface}}$) of 140 meV. As the number of ice bilayers increases, we anticipate this enhanced stability will be proportional to the surface-to-volume ratio. From our previous work[15], we have shown that the bulk limit for the relative energy per water molecule between ice XI and ice Ih is 3.68 meV. The final energy calculation at 0 K is as follows:

$$E^{\mathrm{T=0}}_{\mathrm{Relative/H_2O}} = \frac{\Delta E_{\mathrm{Surface}}}{N_{\mathrm{Bilayers}}} - 3.68\ \mathrm{meV} \tag{S4}$$

We can perform a similar analysis with a Pt slab. As shown in the previous section, we expect the skin layer thickness to be much larger with Pt electrodes, thus the stability of the proton ordered interface should be enhanced by much more than 140 meV. In this case, it is more straightforward to use the fully ferroelectric system shown in Fig.S12a to determine the difference in stability between the ice XI-Au[111] and ice XI-Pt[111] interfaces. Near the center of the ice XI slab, the Pt system is favored by 38 meV. This increases to 205 meV at the interface, implying a surface enhancement of 167 meV. Adding this to the previously discussed enhanced stability in Au of 140 meV, we find that the proton ordered (XI) interface with Pt has an enhanced stability ($\Delta E_{Surface}$) of 307 meV relative to the proton disordered (Ih) interface with Pt.

**B.** Free energy differences versus Temperature.

The treatment of nuclear quantum effects will be summarized here and is described in full detail in our previous work[15]. We define $V_0$ to be the volume that minimizes the Born-Oppenheimer energy, $E_0(V)$. We can expand this energy in a Taylor series for small perturbations from $V_0$.

$$E_0(V) = E_0(V_0) + \frac{B_0}{2V_0}(V - V_0)^2 \tag{S5}$$

The phonon frequencies can also be expanded in this way.

$$\omega_k(V) = \omega(V_0)\left(1 - \gamma_k \frac{V-V_0}{V_0}\right) \tag{S6}$$

In these expansions, $B_0$ is the dominant part of the bulk modulus with vibrational corrections removed. $\gamma_k$ are the mode Grüneisen parameters, defined as

$$\gamma_k = -\frac{\partial(ln\omega_k)}{\partial(lnV)} = -\frac{V}{\omega_k}\frac{\partial\omega_k}{\partial V}. \tag{S7}$$

We calculate the phonon frequencies, $\omega_k$, at three different volumes and calculate its volume dependence to linear order. This adds a volume dependence to the Helmholtz free energy $F(V,T)$[16] of independent harmonic oscillators.

$$F(V,T) = E_0(V) + \sum_k \left[\frac{\hbar\omega_k(V)}{2} + k_B T \ln\left(1 - e^{-\hbar\omega_k(V)/k_B T}\right)\right] - TS_H \tag{S8}$$

The sum is over both phonon branches and phonon wave vectors within the Brillouin zone.

The entropy of the proton disorder, $S_H$, is included in the last term of the free energy equation. This entropy is zero for ice XI (proton-ordered ice). Pauli has estimated that the entropy for ice Ih (proton-disordered ice) is $Nk_B\ln(3/2)$, which was obtained theoretically[17] and experimentally[18,19].

We can find the classical limit of the free energy by taking the high-temperature limit of (S8).

$$\mathrm{F(V,T)} = \mathrm{E_0(V)} + \sum_{\mathrm{k}} \left( \mathrm{k_B T ln} \left\{ \frac{\hbar\omega_{\mathrm{k}}[\mathrm{V(T)}]}{k_B T} \right\} \right) - \mathrm{T}S_H \tag{S9}$$

In Fig 4c of the manuscript, we show the free energies of bulk ice Ih and bulk ice XI as a function of temperature. The difference between these two free energies should match the enhanced stability described in the previous section at the transition temperature.

$$\Delta F_{(\mathrm{XI-Ih})/\mathrm{H_2O}}\,(\mathrm{T_C}) \;=\; -\left(E^{\mathrm{T=0}}_{\mathrm{Relative/H_2O}} + 3.68\ \mathrm{meV}\right) \tag{S10}$$

The skin layer thickness, $h$, can be found from the number of bilayers with the conversion factor 0.3615 nm / 1 bilayer. After combining (S4) and (S10), we are left with the following equation:

$$h\,(T_C) \;=\; \frac{\Delta E_{\mathrm{Surface}} \cdot 0.3615\ \mathrm{nm}}{\Delta F_{(\mathrm{XI-Ih})/\mathrm{H_2O}}\,(T_C)} \tag{S11}$$

Substituting the experimentally obtained phase-transition temperatures into equation (S11), we find $h$ to be 14.7 nm for Au and 34.7 nm for Pt, in reasonable agreement with the experimentally estimated $h$ of 20.3nm for Au and 31.3nm for Pt in Section4.

**S6. The source of free charge and its characteristic time**

The most promising candidate of the transferred charge species in ice charging problem is considered to be the mobile ions ($H^+$, $OH^-$, and other impurities) in the quasi-liquid layers (QLL) on ice surfaces[20-23]. The existence of a QLL with thickness of 1~100 nm on the surface of ice has been verified by many techniques[24-29] and summarized by authoritative review articles[20,30,31]. Our own data also indicates the appearance of QLL above 248K (Fig. S4). Below we list some consequences of the existence of QLL layers on the electrical properties of ice surface.

The ionic charge density in QLL has been determined to be in the range of $10^{-3}$~$10^{-2}$ C/m$^2$ by synchrotron X-radiation[32], scanning force microscopy[33], and molecular dynamic calculations[34]. This charge density is sufficient to compensate for our calculated flexoelectric polarization ($10^{-5}$~$10^{-4}$ C/m$^2$). In addition, a volta-effect measurement[35] confirms that the appearance of the QLL above 243K alters the ice surface potential by ~100mV, which indicates the change in charge density at the surface and is indeed comparable with the zeta potential of the ice/water interface[36,37]. The mobility of surface charge carriers of ice is reported, despite the large discrepancies between Hall-effect measurements[38] (~$3\times10^{-4}$ m$^2$/Vs) and field effect transistor measurements[39] ($\sim 3\times10^{-7}$ $m^2/Vs$), to be larger than that in bulk ice. Also, direct electrical measurements[39-41] show the surface conductance of ice to be $10^{-11}\sim10^{-9}$ $\Omega^{-1}$. Assuming a moderate QLL thickness of 10nm[31,42], the surface conductivity of ice can be estimated to be $10^{-3}\sim10^{-1}$ S/m, orders of magnitude higher than that of bulk ice[43]. The abundant literature on the matter thus provides overwhelming evidence that ice particles are coated with a QLL that contains mobile charges capable of responding to electric fields.

The next question is whether the electrical relaxation time $\tau_e$ of QLL is comparable with mechanical contact time, $\tau_m$. In particular, if $\tau_e > \tau_m$, the free charge will not have time to flow to the contact area within the timescale of the induced deformation, so for our model to be feasible it is necessary that $\tau_e \leq \tau_m$.

Typical $\tau_m$ is in the range of $10^{-7}\sim10^{-5}$ s, depending on the particle size and impact velocities[44,45], which can also be directly calculated from Hertzian mechanics[46]. Meanwhile, $\tau_e$ can be estimated as the Maxwell relaxation time[47] (RC constant) $\tau_e = \xi_r \xi_0/\sigma$, where $\xi_r$ and $\sigma$ are the dielectric constant and surface conductivity of ice. Adopting $\xi_r$ of 100 (Table S1) and $\sigma$ of $10^{-3}\sim10^{-1}$ S/m (last paragraph), we can estimate $\tau_e$ in a range of $10^{-8}\sim10^{-6}$ s. This, of course, is a back-of-envelope estimation. A more rigorous theoretical treatment[48] estimated a characterization time of 1.4 μs. The condition $\tau_e \leq \tau_m$ is therefore met, indicating the plausibility of QLL as a source for screening charge. However, we cannot exclude other possible sources of charge species (see S11). More generally, the nature and origin of charge species remains an unresolved issue in the triboelectric community[49,50].

**S7. Alternative explanations for ice charging mechanisms**

1) <u>Workman-Reynolds freezing potential.</u> The process of freezing dilute aqueous solutions can lead to charge separation due to unequaled incorporation of solute ions into ice lattice, suggested to contribute to the contact electrification (CE) between ice particles and supercooled water droplets[51,52]. Such freezing potential, however, is found to develop very slowly and cannot explain the significant CE with short contact time in thunderstorms[53,54].

2) <u>External electric field.</u> Although the presence of an external electric field facilitates the CE of ice[55,56], there is plenty of evidence showing that the CE of ice can also occur in the absence of such a field[45,57-65]. Also, invoking external electric fields requires answering the question of where they come from.

3) <u>Temperature difference.</u> The thermoelectric effect has been used to explain the CE of ice under a temperature difference $\Delta T$ between two contact bodies[66-68]. However, many studies have reported results that contradict this explanation, including the observed CE of ice in the absence of $\Delta T$ or the observed CE in the presence of $\Delta T$ but with a charge-transfer direction opposite to the theoretical prediction[57,58,62,64].

4) <u>Surface ion concentration difference.</u> This is the most popular explanation so far for the CE of ice[20,53,65]. It suggests that if ice and graupel particles are both grown from vapor and with different growth rates, the two surfaces would process different charge densities in the QLL that drives ion diffusion (transfer) during contact[20,53,65,69-71]. However, the CE of ice has also been observed when both ice and graupel were grown from freezing droplets instead of vapor[57,61,62,64,72]. In line with this, this explanation has clear shortcomings, and C.P.R. Saunders, one of its inventors[69], commented[65] "*it offers no ability to quantitatively predict the magnitude of charge transferred during ice particle collisions, nor insight into explicit physical, microscopic charge transfer processes*".

5) <u>Mechanochemistry.</u> It is not surprising that the contact between two surfaces can involve the breaking/formation of chemical bonds, exposure of new surfaces, and exchange of mass. These processes, when equipped with a certain difference between two contact bodies, have been proposed to explain ice charging. For example, fractures + temperature difference[67,73], contact sintering + surface ion concentration difference[44], collisional melting + mass transfer + surface ion concentration difference[23,74], pressure melting + mass transfer + geometric shape difference[75,76]. However, it is challenging to quantitatively evaluate the role of mechanochemistry in the CE, given the enormous uncertainties involved.

A more complete list can be found in review articles[20,53]. The common message of these studies is that the understanding of the CE of ice is complex and far from complete. It can involve different physical mechanisms that can co-exist and interact in subtle ways, which may explain why it remains unsolved till now as one of the earliest scientific problems of all time. As C.P.R. Saunders, an expert intensively working on the ice charging problem[53,58,69], commented recently[71] that "*meaningful forecasts of storm electrical evolution and lightning properties are not currently possible because there remains poor understanding of the fundamental physics of the underlying mechanisms that account for cloud charging and the complex charge distributions observed.*" Rohan Jayaratne, another expert in the field[58,69], also pointed out that very recently[44] that "*these various hypotheses have been found to be inconsistent with the experimental evidence when subjected to careful scrutiny, and a new direction of approach to understanding the microphysical charging mechanism is necessary*".

It is not our intention to claim ice flexoelectricity as *the* new direction of approach that will put an end to such a long-standing debate, but to show that its contribution to the CE of ice cannot be ignored, since it is conceptually inevitable (flexoelectricity is generated in any inhomogeneous deformation), and quantitatively relevant.

## S8. Calculation of flexoelectric polarization and field

Flexoelectric polarization is defined as[77,78]

$$P_i = \mu_{ijkl}\frac{\partial \varepsilon_{kl}}{\partial x_j} \quad \text{(S12)}$$

where $\mu_{ijkl}$ is the fourth-rank tensor of flexoelectric coefficients, $P_i$ is the electric polarization, $\varepsilon_{kl}$ is the strain tensor, and $x_j$ is the position coordinate. We consider three components of $\mu_{ijkl}$ (longitudinal $\mu_{1111}$, transversal $\mu_{1122}$, and shear $\mu_{1212}$) in our calculations, which can be rewritten in the Voigt notation as $\mu_{11}$, $\mu_{12}$, and $\mu_{44}$, respectively. The polarization field in Cartesian coordinates can be written as[79,80]

$$P_x = \mu_{11}\frac{\partial \varepsilon_{xx}}{\partial x} + \mu_{12}(\frac{\partial \varepsilon_{yy}}{\partial x} + \frac{\partial \varepsilon_{zz}}{\partial x}) + 2\mu_{44}(\frac{\partial \varepsilon_{xy}}{\partial y} + \frac{\partial \varepsilon_{xz}}{\partial z})$$

$$P_y = \mu_{11}\frac{\partial \varepsilon_{yy}}{\partial y} + \mu_{12}(\frac{\partial \varepsilon_{xx}}{\partial y} + \frac{\partial \varepsilon_{zz}}{\partial y}) + 2\mu_{44}(\frac{\partial \varepsilon_{xy}}{\partial x} + \frac{\partial \varepsilon_{yz}}{\partial z}) \quad \text{(S13)}$$

$$P_z = \mu_{11}\frac{\partial \varepsilon_{zz}}{\partial z} + \mu_{12}(\frac{\partial \varepsilon_{xx}}{\partial z} + \frac{\partial \varepsilon_{yy}}{\partial z}) + 2\mu_{44}(\frac{\partial \varepsilon_{xz}}{\partial x} + \frac{\partial \varepsilon_{yz}}{\partial y})$$

Since it is not possible to measure separately each of the tensor components, assumptions are needed to determine these components from the measured $\mu_{13}^{eff}$ and their relationships. The $\mu_{13}^{eff}$ measured in a standard bending method is linked to $\mu_{11}$ and $\mu_{12}$ by[77] $\mu_{13}^{eff} = -\nu\mu_{11} + (1-\nu)\,\mu_{12}$, where $\nu$ is the Poisson's ratio (0.325, Table S1). In parallel, for isotropic solids, $\mu_{44}$ is linked to $\mu_{11}$ and $\mu_{12}$ by[81] $\mu_{12} = (\mu_{11} - \mu_{44})/2$. Assuming $\mu_{11} = \mu_{12}$, we have $\mu_{11} = \mu_{12} = -\mu_{44} = \mu_{13}^{eff}/(1-2\nu) = 2.86\mu_{13}^{eff}$. Then we can obtain the flexoelectric polarization from equation (S13) using the calculated strain gradients (see the next section). Flexoelectric polarization creates depolarization field that opposes to it, which in isotropic solids (relative permittivity $\xi_r$ is a constant) is given by[77,78]

$$E_x = -P_x/\xi_0\xi_r, \quad E_y = -P_y/\xi_0\xi_r, \quad E_z = -P_z/\xi_0\xi_r \quad \text{(S14)}$$

## S9. Calculation of flexoelectricity in ice-graupel collisions

The classical Hertzian theory of contact mechanics has been widely used to estimate the strain distributions and the associated flexoelectricity in contact problems[82-84]. The Hertzian solution of static contact can also apply to impact problem of two elastic spheres with each body treated as an elastic half-space[46]. The maximum effective impact force is given by[46]

$$F^* = \frac{4}{3} Y^* \sqrt{R^*} \left( \frac{15 m^* {v_r}^2}{16 Y^* \sqrt{R^*}} \right)^{3/5} \tag{S15}$$

where $Y^*$, $R^*$, and $m^*$ are the effective Young's modulus, effective radius, effective mass of the impact, $v_r$ is the relative velocity of two contact bodies (ice and graupel particle). These effective properties are given by[85]

$$\frac{1}{Y^*} = \frac{1-{\nu_i}^2}{Y_i} + \frac{1-{\nu_g}^2}{Y_g},\ \frac{1}{R^*} = \frac{1}{R_i} + \frac{1}{R_g},\ \frac{1}{m^*} = \frac{1}{m_i} + \frac{1}{m_g} \tag{S16}$$

where $Y_i$, and $Y_g$, $R_i$ and $R_g$, $m_i$ and $m_g$, $\nu_i$ and $\nu_g$ are Young's modulus, radius, mass, and Poisson's ratio of the ice and graupel particles respectively. $Y_i$ and $\nu_i$ are easily accessible in the literature[43] (Table S1). $Y_g$ and $\nu_g$ are less so. Given graupel particles are most parameterized as "medium density" of 400~500 kg/m$^3$ [53,69,86-89], and based on the density-modulus relationship[43], we assume a representative modulus for $Y_g$ of 1 GPa (Table S1). Still, the dependence of the results on $Y_g$ has also been studied as we will see later (Fig. S13d). On the other hand, we could not find a representative value for $\nu_g$, so we assume $\nu_g = \nu_i = \nu$. In parallel, considering that $R_i$ is often orders of magnitude smaller than $R_g$[43,53], the last two formulas in (S16) can be simplified as $R^* = R_i$ and $m^* = m_i = \frac{4}{3}\rho\pi {R_i}^3$, where $\rho$ is the density of ice (Table S1).

Under load $F^*$, the contact radius $a$ and mean pressure $P_m$ on the contact surface can be expressed as[85]

$$a = \sqrt[3]{\frac{3}{4}\frac{F^* R^*}{Y^*}},\quad P_m = \frac{F^*}{\pi a^2} \tag{S17}$$

The stress fields of the semi-space beneath the contact region of a spherical indenter in cylindrical coordinates are given by[85]

$$\sigma_{rr} = \frac{3P_m}{2}\left\{\frac{1-2\nu}{3}\frac{a^2}{r^2}\left[1 - \left(\frac{z}{\sqrt{u}}\right)^3\right] + \left(\frac{z}{\sqrt{u}}\right)^3 \frac{a^2 u}{u^2+a^2z^2} + \frac{z}{\sqrt{u}}\left[u\frac{1-\nu}{a^2+u} + (1+\nu)\frac{\sqrt{u}}{a}tan^{-1}\frac{a}{\sqrt{u}} - 2\right]\right\} \tag{S18}$$

$$\sigma_{\theta\theta} = -\frac{3P_m}{2}\left\{\frac{1-2\nu}{3}\frac{a^2}{r^2}\left[1 - \left(\frac{z}{\sqrt{u}}\right)^3\right] + \frac{z}{\sqrt{u}}\left(2\nu + u\frac{1-\nu}{a^2+u} - (1+\nu)\frac{\sqrt{u}}{a}tan^{-1}\frac{a}{\sqrt{u}}\right)\right\} \tag{S19}$$

$$\sigma_{zz} = -\frac{3P_m}{2}\left\{\left(\frac{z}{\sqrt{u}}\right)^3 \frac{a^2 u}{u^2+a^2z^2}\right\} \tag{S20}$$

$$\sigma_{rz} = -\frac{3P_m}{2}\left(\frac{rz^2}{u^2+a^2z^2}\right)\left(\frac{a^2\sqrt{u}}{a^2+u}\right) \tag{S21}$$

where $u$ is the displacement of points on the contact surface, and can be expressed as

$$u = \frac{1}{2}\left[(r^2 + z^2 - a^2) + \sqrt{(r^2 + z^2 - a^2)^2 + 4a^2z^2}\right] \tag{S22}$$

Note that the above stress fields are symmetric in two contact bodies, but strain fields are not because of their different Young's modulus. The strain fields beneath the contact region in cylindrical coordinates can be expressed by the isotropic Hooke's law

$$\begin{aligned} \varepsilon_{rr}^{g} &= \frac{1}{Y_g}\left[\sigma_{rr} - \nu(\sigma_{\theta\theta} + \sigma_{zz})\right], \; \varepsilon_{rr}^{i} = \frac{Y_g}{Y_i}\varepsilon_{rr}^{g} \\ \varepsilon_{\theta\theta}^{g} &= \frac{1}{Y_g}\left[\sigma_{yy} - \nu(\sigma_{rr} + \sigma_{zz})\right], \quad \varepsilon_{\theta\theta}^{i} = \frac{Y_g}{Y_i}\varepsilon_{\theta\theta}^{g} \\ \varepsilon_{zz}^{g} &= \frac{1}{Y_g}\left[\sigma_{zz} - \nu(\sigma_{rr} + \sigma_{\theta\theta})\right], \quad \varepsilon_{zz}^{i} = \frac{Y_g}{Y_i}\varepsilon_{zz}^{g} \\ \varepsilon_{rz}^{g} &= \frac{1+\nu}{Y_g}\sigma_{rz}, \quad \varepsilon_{rz}^{i} = \frac{Y_g}{Y_i}\varepsilon_{rz}^{g} \end{aligned} \tag{S23}$$

According to equation (S13) and the axial symmetry in this problem, the vertical polarization beneath the contact region can be written in cylindrical coordinates as

$$P_z^g = \mu_{11}\frac{\partial \varepsilon_{zz}^g}{\partial z} + \mu_{12}\left(\frac{\partial \varepsilon_{rr}^g}{\partial z} + \frac{\partial \varepsilon_{\theta\theta}^g}{\partial z}\right) + 2\mu_{44}\frac{\partial \varepsilon_{rz}^g}{\partial r}, \; P_z^i = \frac{Y_g}{Y_i}P_z^g \tag{S24}$$

Combining equations (S15) to (S24) and adopting typical parameters in ice-graupel collisions (Table S1), we can compute the distributions of the flexoelectric polarization beneath the contact interface in two bodies analytically. Because the graupel particle is softer, it experiences larger deformation and generates larger polarization ($P_z^g$) than that of ice particle ($P_z^i$). Fig. 5a shows the flexoelectric polarization in a specific case with $v$=8 m/s and $R$=50 $\mu$m, with the corresponding depolarization field (using equation (S14)) shown in Fig. S13a. Note that Hertzian analytic solutions are not adequate to show the actual deformation, we instead use finite element simulation (see next section for details) to plot Fig. 5a and Fig. S13a for schematic illustration. Integrating the net interfacial polarization ($P_z = P_z^g - P_z^i$) over the contact interface, we can estimate the upper bound of flexoelectric-driven transferred charge

$$Q = \int_0^a P_z 2\pi r \, dr = \left(1 - \frac{Y_g}{Y_i}\right)\int_0^a P_z^g 2\pi r \, dr \tag{S25}$$

We can further rewrite this equation in a more explicit form based on dimensional analysis. The stress $\sigma_{ij}$ in equations (S18)~(S21) can be expressed by a change of variables as $P_m\phi_{ij}(\frac{r}{a},\frac{z}{a},\nu)$, where $\nu$ is Poisson's ratio and $\phi_{ij}$ is a dimensionless function. The strain of graupel $\varepsilon_{ij}^g$ in equation (S23) can then be expressed as $\frac{P_m}{Y_g}\psi_{ij}(\frac{r}{a},\frac{z}{a},\nu)$, where $\psi_{ij}$ is another dimensionless function. The polarization of graupel $P_z^g$ in equation (S23) inherits the same scaling of $\varepsilon_{ij}^g$ but acquires additional factors of $\mu_{13}^{eff}$ (see section 8 for the relationship between $\mu_{13}^{eff}$ and $\mu_{ij}$) and $\frac{1}{a}$ through differential. With this, and setting $z = 0$ (at the interface), we can write $P_z^g$ as $\frac{\mu_{13}^{\mathrm{eff}} P_m}{a Y_g} \varphi(\frac{r}{a}, \nu)$, where $\varphi$ is another dimensionless function. Let $t = \frac{r}{a}$, the integral part in equation (S25) can be written as

$$\int_0^a P_z^g 2\pi r\, dr = \frac{\mu_{13}^{\mathrm{eff}} a P_m}{Y_g} \alpha(\nu) \tag{S26}$$

where the defined function $\alpha(\nu)$ is equal to $2\pi \int_0^1 t\varphi(t, \nu)\, dt$. This term absorbs the surface integration, only depends on Poisson's ratio, and remains the only term that requires numerical calculations. Substituting equation (S15)~(S17) into equation (S26), we have

$$\int_0^a P_z^g 2\pi r\, dr \approx 0.73352\, \mu_{13}^{\mathrm{eff}} R_i {v_r}^{\frac{4}{5}} \rho^{\frac{2}{5}} \frac{{Y_i}^{\frac{3}{5}} {Y_g}^{-\frac{2}{5}}}{(Y_i + Y_g)^{\frac{3}{5}}} \frac{\alpha(\nu)}{(1-\nu^2)^{\frac{3}{5}}} \tag{S27}$$

We plot the value of $\alpha(\nu)$ against $\nu$ in Fig. S13b. In the specific case here in which $\nu = 0.325$, we have $\alpha(0.325) \approx 19.744$. Substituting this value into equation (S27) and then back to equation (S25), we obtain

$$Q \approx 15.486 \mu_{13}^{\mathrm{eff}} R_i {v_r}^{\frac{4}{5}} \rho^{\frac{2}{5}} \frac{Y_i - Y_g}{(Y_i + Y_g)^{\frac{3}{5}} (Y_i Y_g)^{\frac{2}{5}}} \tag{S28}$$

This equation shows that $Q = 0$ when $Y_i = Y_g$: the existence of nonzero flexoelectric charge (in Hertzian contact approximation) hinges on the Young's modulus of the graupel and ice particle being unequal, as illustrated by the plot of $Q$ against $Y_g/Y_i$ in Fig. S13c. Equations (S25) ~ (S28) should be applicable to the contact charging in other material systems. When doing so, please note that the current formulation assumes identical Poisson's ratio $\nu$ and effective flexoelectric coefficient $\mu_{13}^{\mathrm{eff}}$ for two contact bodies. Modifications may be required, particularly when dealing with chemically different materials (e.g., $\mu_{13}^{\mathrm{eff}} = 0$ for metals).

**S10. Details of the comparison between theory and literature experiments**

We compare the calculated $Q$ with some experimentally reported $Q$ in each ice-graupel collision event[57-61,63,64]. Given that the used $R_i$ and $v_r$ often varies simultaneously among these studies, and considering the proportional relationship of $Q \propto R_i {v_r}^{\frac{4}{5}}$ indicated by equation (S27), we look at the dependence of $Q$ on $R_i {v_r}^{\frac{4}{5}}$ in the comparison. The temperature range in these experiments[57-61,63,64] are generally within 243K~273K, with graupels being negatively charged at colder regions while positively charged at warmer regions, the specific temperatures used were different. Considering this, and the electrode-sensitive anomaly of flexoelectricity near the melting point, we chose to use a conservative $\mu_{13}^{eff}$ (1.14±0.13 nC/m), the intrinsic value obtained between 203~248K, in the theory and compare the theoretical result with the experimental $Q$ in the negative charging regime for graupels. This is also consistent with the schematic illustration in Fig. 5a. Below we list the details of extracting the literature data. Note that our wording from now on no longer emphasize the sign of $Q$ but refer only to its absolute values, considering the same polarity shared by these data, i.e., negative charge acquired by graupels. Also note that some of these experiments were performed under different environmental water content that also affects the $Q$, but we currently are not able to involve this extra variable in our analysis.

1. 1980 Gaskell et al[57], represented by purple triangle symbols in Fig. 5c. Gaskell systematically studied the $v_r$ dependence of $Q$ when $R_i = 50\mu m$ and $R_i$ dependence of $Q$ when $v_r = 8m/s$ in the negative charging regime of graupels. We extract the first dataset from Figure 3 that showed "*histograms of the charges transferred to the hailstone for each velocity of impact of the 100μm diameter ice spheres*" and obtained the weighed mean and standard deviation of $Q$ under different $v_r$ but the same $R_i$. We extract the second dataset of $R_i$ and $Q$ from Figure 4 that showed "*average charge (Q) transferred to the hailstone for each diameter of the ice spheres colliding at a velocity of 8 m s$^{-1}$*" and obtained the mean of $Q$ under different $R_i$ but the same $v_r$. The uncertainty of the second dataset was not accessible in Gaskell's paper, but instead in Figure 6.7 of Gaskell's thesis[90].
2. 1983 Jayaratne et al[58], represented by orange sphere symbols in Fig. 5c. We took two datasets. In the first one, $R_i$, $v_r$ and $Q$ are all described in this sentence on page 621: "*At a flow speed of 9.8ms$^{-1}$ with the target and cloud in thermal equilibrium, the mean charge acquired by the target per rebounding crystal of size 30μm at a temperature of about -10* ℃*, assuming an event probability of 0.2, was -0.25 fC.*" The second dataset was the negative rime charging datapoints at -21℃ in Figure 8, which reported the $R_i$ dependence of $Q$ when $v_r$ =2.8 m/s. Note that the data points in Figure 8 were revised by the same author, based on more accurate estimations of collision probability, in a later paper[91], from which we extract the revised data points (solid sphere symbols in Figure 13a).
3. 1989 Keith et al[59], represented by green triangle symbols in Fig. 5c. The experimental $v_r$, $R_i$ and $Q$ are described in the following words on page 57: "*The rod speed was 3.2 m s$^{-1}$ which is consistent with that of previous workers and with the fall speed of a hailstone in a thunder cloud*"; "*The charge per crystal separation event at the reference stage, as denoted by the arrow, is 1.8 fC for 90 μm diameter crystals and a liquid water content of 0.30 gm$^{-3}$*"; "*The charge per event at the reference stage for a 100 um crystal was 1.9 fC for a liquid* *water content of 0.5 gm$^{-3}$.*"; "*The charge per event at the reference stage for a 65 μm crystal was 0.3 fC for a liquid water content of 0.6 gm$^{-3}$.*"

4. 1991 Caranti et al[60], represented by yellow triangle symbols in Fig. 5c. The experimental $R_i$ and $v_r$ are described in this sentence on page 366: "*Ice particles of 100μm in diameter and impact velocities between 6 and 10 m/s were used*". The extraction of $Q$ must be very careful because this paper investigated the dependence of $Q$ on the temperature difference $\Delta T$ between ice and graupel (see #3 in section7), manifested in the significantly high $Q$ in the bottom graph in Figure 2 when $\Delta T$=7°C. Such an additional variable is absent in other literature experiments and in our model, we therefore chose to only extract the histogram data corresponding to $\Delta T = 0$, i.e., the top graph in Fig.2. After excluding two extreme values at both ends with low probability, we obtained the weighted mean of 23.03 fC and standard deviation of 19.38 fC.
5. 1994 Avila et al[61], represented by red square symbols in Fig. 5c. The experimental $R_i$ and $v_r$ are described in the abstract: "*A cylinder growing by riming in a wind tunnel was used as target for collisions at 5 m s$^{-1}$ with ice spheres of 100 μm in diameter in a wide range of temperatures.*" In the reported charge histograms in Figure 3 and Figure 4, only the first graph (at -24°C) in Figure 3 corresponds to the negative charging of graupel. After excluding two extreme values at both ends with low probability, we obtained the weighted mean of 20.64 fC and standard deviation of 17.90 fC.
6. 2002 Pereyra et al[64], represented by blue diamond symbols in Fig. 5c. The experimental $v_r$ are described in the abstract: "*The measurements were performed with an impact velocity of 8.5 m s$^{-1}$*". Also in the abstract, the author found that "*the artificial graupel charges positively for temperatures above -12°C and negatively for temperatures below -14°C.*" We therefore extract the $R_i$ from the bottom graph (*-20°C~-15°C*) of Figure 3, finding a weighed mean of 13.41 μm and standard deviation of 6.61 μm. By the same token, we extract the $Q$ from the only five graphs in Figure 5 that were measured below -14°C, finding a weighed mean of 8.79 fC and standard deviation of 21.58 fC.
7. 2020 Luque et al[63], represented by the grey shadow in Fig. 5c. The authors put that "*The simulated graupel charges negatively for temperatures below −15 °C, and positively at temperatures above −15 °C.*" and "*the average charge transferred per collision between −1 and −3 fC in the negative charging regime*". The experimental $v_r$ was described in the abstract: "*ambient temperatures between −5 and −30 °C and at air speed of 11 m s$^{-1}$*". The experimental $R_i$ was listed by six histograms in their Fig. 2. We added up four $R_i$ panels in the negative charging regime (−17°C, −21°C, −26°C, −30°C), finding a weighed mean of 12.33 $\mu$m and a standard deviation of 5.35 $\mu$m.

**S11. Finite element simulation**

To show the actual deformation of ice-graupel collision in Fig. 5b and Fig.13a, the finite element simulation is performed in COMSOL Multiphysics 6.0 in the Solid Mechanics module. Linear elasticity is assumed, and axisymmetric configuration is adopted. Material, geometric, and load parameters are the same as the previous section. As shown in Fig. 14, the ice particle is modelled as a half sphere and the graupel particle is modelled as a cuboid to represent an effectively infinite radius of curvature. We define the ice–graupel interface as a contact pair, with the graupel's top surface set as the destination and the ice's bottom surface as the source. A total force $F^*$, from equation (S15), is applied on the ice's upper surface, where a spring foundation is also imposed for helping convergence. The augmented Lagrangian approach with separated solutions is used to model contact, with an initial contact pressure of 10 MPa. Refined quadrilateral elements are used near the contact zone, while the remainder in the model was discretized using coarser triangular elements. On top of this, the destination boundary (graupel surface) is meshed finer than the source boundary (ice surface) by a factor of two. The calculated strain fields are then fed to equation (S24) for the polarization fields (Fig. 5b) and then to equation (S14) for depolarization fields (Fig.13a).

**S12. Some things we do not know**

Flexoelectricity, ice (surface), and triboelectricity are already complex subjects in their own, and their combination only exacerbates the complexity. Below we list some of the simplifications and neglected factors we have assumed to make our formulation tractable within the scope of this article. Each of them may affect the electrical and mechanical dynamics during the contact and eventually amplify (or diminish) the final share of flexoelectricity. Cooperative efforts from different communities will likely be needed to complete the picture and fully determine the role of flexoelectricity in the ice charging problem.

1. Linear elasticity is assumed. In reality, the collision may involve anelastic losses, plasticity and fractures, and new defects can be generated. See references[92-94] for how dislocations and fractures influence flexoelectricity.
2. Perfect sphere shapes and smooth surfaces are assumed. In reality, ice and graupel particles do not have to be spherical and certainly contain asperities on surfaces. See references[95,96] for how indenter shape and surface roughness influence flexoelectricity.

3. The graupel particle is treated as continuum body. In reality, graupel particles are porous and have complex microstructures. See references[97-101] for how porous microstructures influence flexoelectricity.
4. Collinear impact and frictionless contact are assumed. In reality, oblique impact can take place and friction can be involved. See references[79,102,103] for how friction affects flexoelectricity and vice versa[104].
5. The flexoelectric coefficient was measured with electrodes that are not involved in ice-graupel collisions. Its potential influence is minimized by adopting the most conservative flexoelectric coefficient ($1.14 \pm 0.13$ nC/m) in our calculation. Future development in electrode-free techniques for quantifying flexoelectricity may provide a more accurate reference.
6. The flexoelectric coefficient was measured at a quasi-static regime (10~17 Hz), but the ice-graupel collision time is typically $10^{-5}$~$10^{-7}$ s. High frequencies, unfortunately, are unavailable in standard bending setups (including ours) for measuring flexoelectricity, limited by the mechanical resonances within DMA. Such resonance limitations may be mitigated when experiments are performed in actuating mode (inverse/converse flexoelectricity). See references[105-109] for how inverse/converse flexoelectricity may be measured.
7. The components of flexoelectric tensors are not individually resolved yet. The effective flexoelectric coefficient we measured is necessarily a combination of different components. Resolving individual tensor components remains a long-standing challenge in the community. Still, future experiments on single crystals with different orientations[7,8,110] and/or experiments in truncated pyramid configuration[111,112] may help, although it has to be noted that deformation experiments alone have been mathematically proven to be incapable of providing the full independent set of tensorial components[110].
8. Hydrodynamics of quasi-liquid layers (QLL) is not considered. The QLL on ice surfaces is ultra-viscous, especially within the inner layer[113,114], and even non-Newtonian under certain conditions[115,116]. It may deform or rupture during collisions and affect (presumably lower) the contact dynamics. Fluid dynamics is already complex on its own, the dynamics of a QLL with a gradient of viscosity are beyond our capabilities – or even comprehension.
9. The origin and nature of the charge species are not determined. Although QLL is considered as the most promising candidate to provide transferred charge (Section 6), it is not an unequivocal conclusion, and other sources may also be at play, such as the pressure-melted liquid (discussed next) and protonic defects in ice lattices[44].

10. Contact-induced phase transition is not considered. The contact pressure on the surface in Fig. 5b is 80 MPa, which in theory reduces the melting point by 6K at ambient pressure. Larger kinetic energy would have a bigger effect. There are debates on whether such process practically occur and contribute to ice charging[20,117,118]. There are also opposing theory arguing instead sintering of QLL on ice surface that participates in ice charging[44]. We have no clue how to examine the influence.
11. Impurities are not considered. Ice in nature is not pure, and impurity ions are known to modify the surface properties of ice[40,119] and (perhaps consequently) affect the contact electrification of ice[58]. In addition, flexoelectricity is also dependent on impurity levels[14,120].

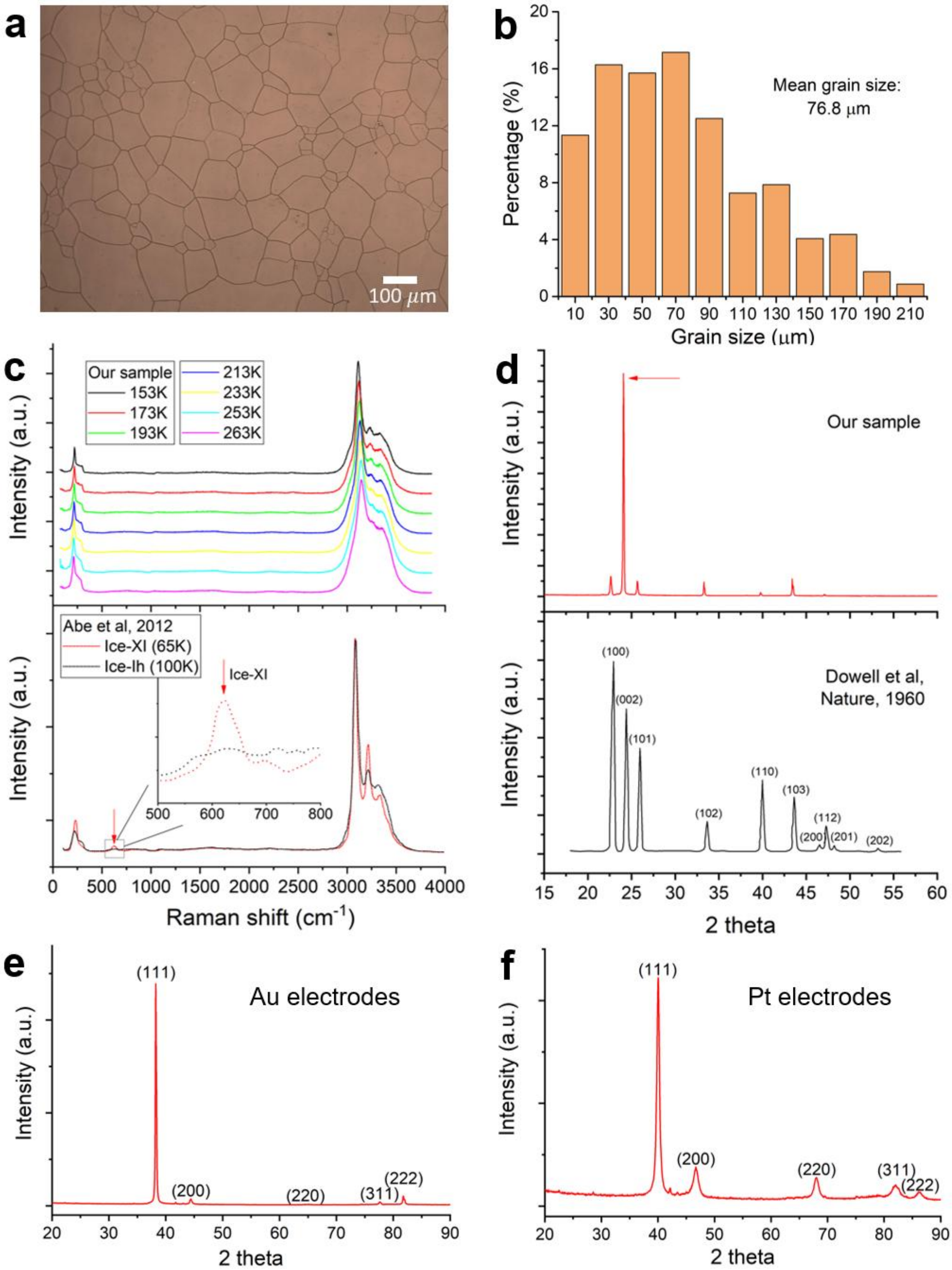

**Fig. S1.** Structural characterizations of our ice sample. **a**, Optical image of the sample surface. **b**, The statistical distribution of the grain size. **c**, The Raman spectrum measured at different temperatures, compared with the reported Raman spectrum for ice-Ih and ice-XI. **d**, The X-ray diffraction pattern measured at 256 K, compared with the reported Xrd pattern for ice-Ih[121]**. e and f,** The X-ray diffraction pattern of Au and Pt electrodes respectively.

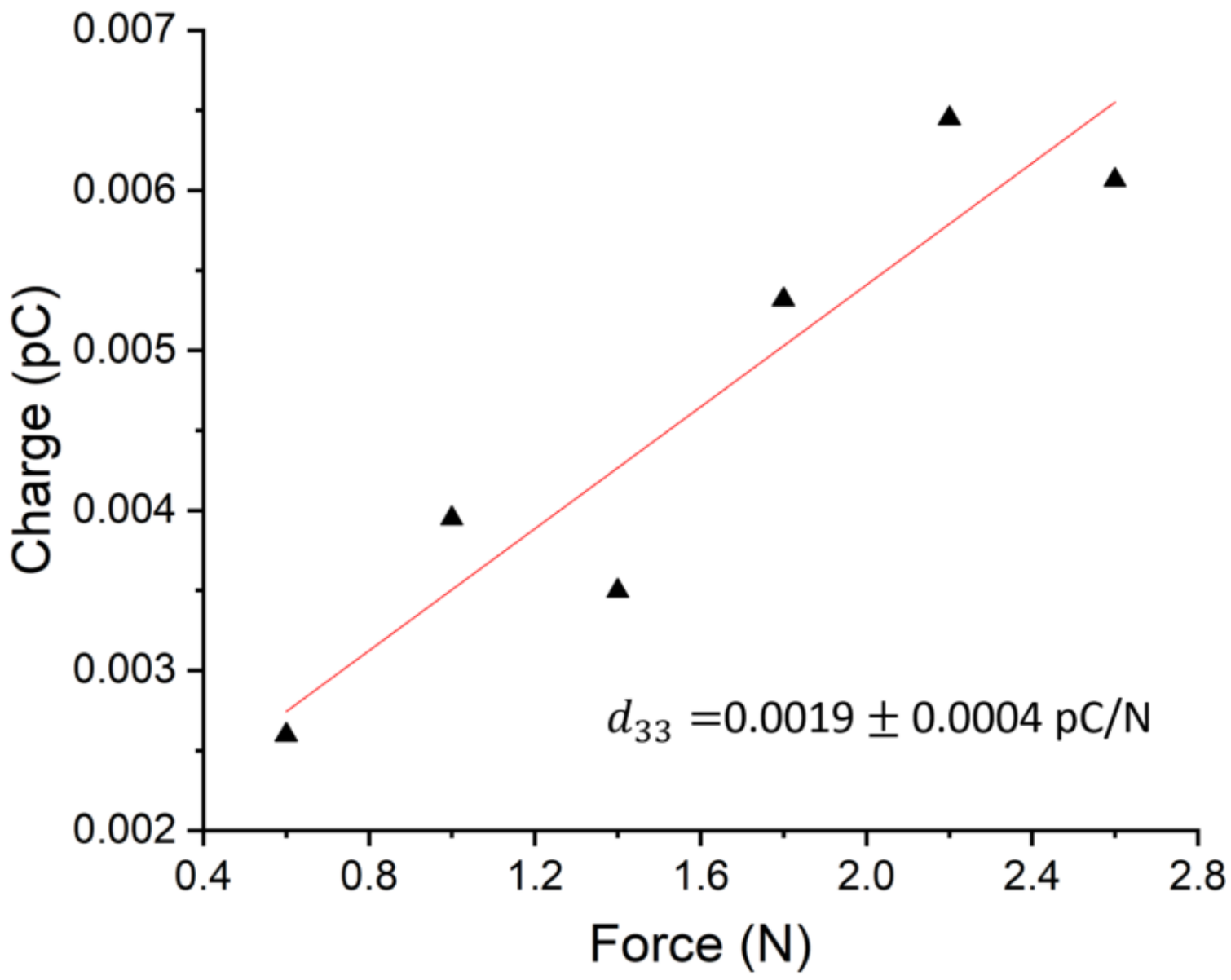

**Fig. S2**. Polarization charge versus the applied force in uniaxial compression experiments measured at ~233 K

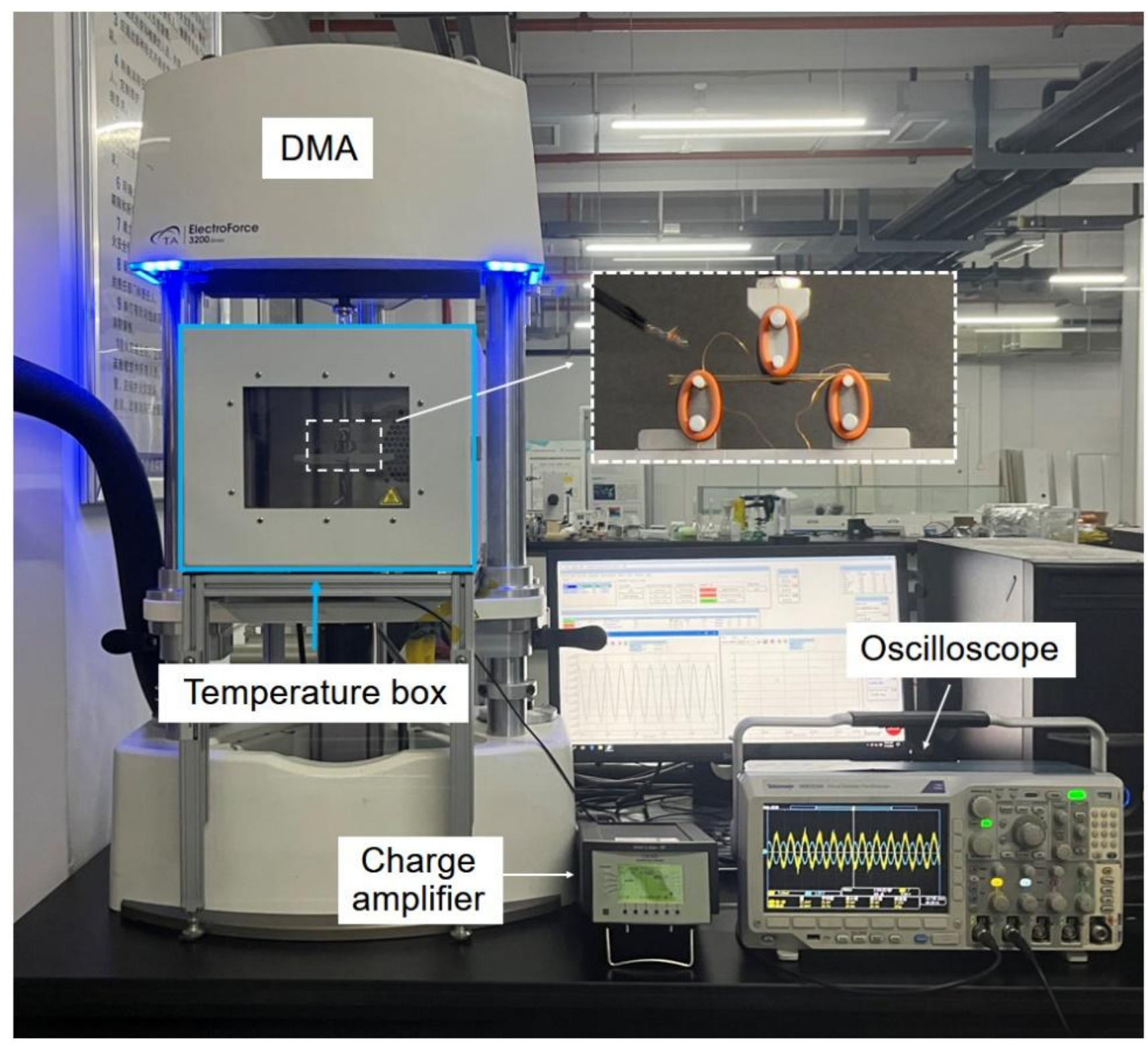

**Fig. S3.** Experimental setup, consisting of a dynamic mechanical analyzer (DMA) with temperature function, a charge amplifier, and an oscilloscope

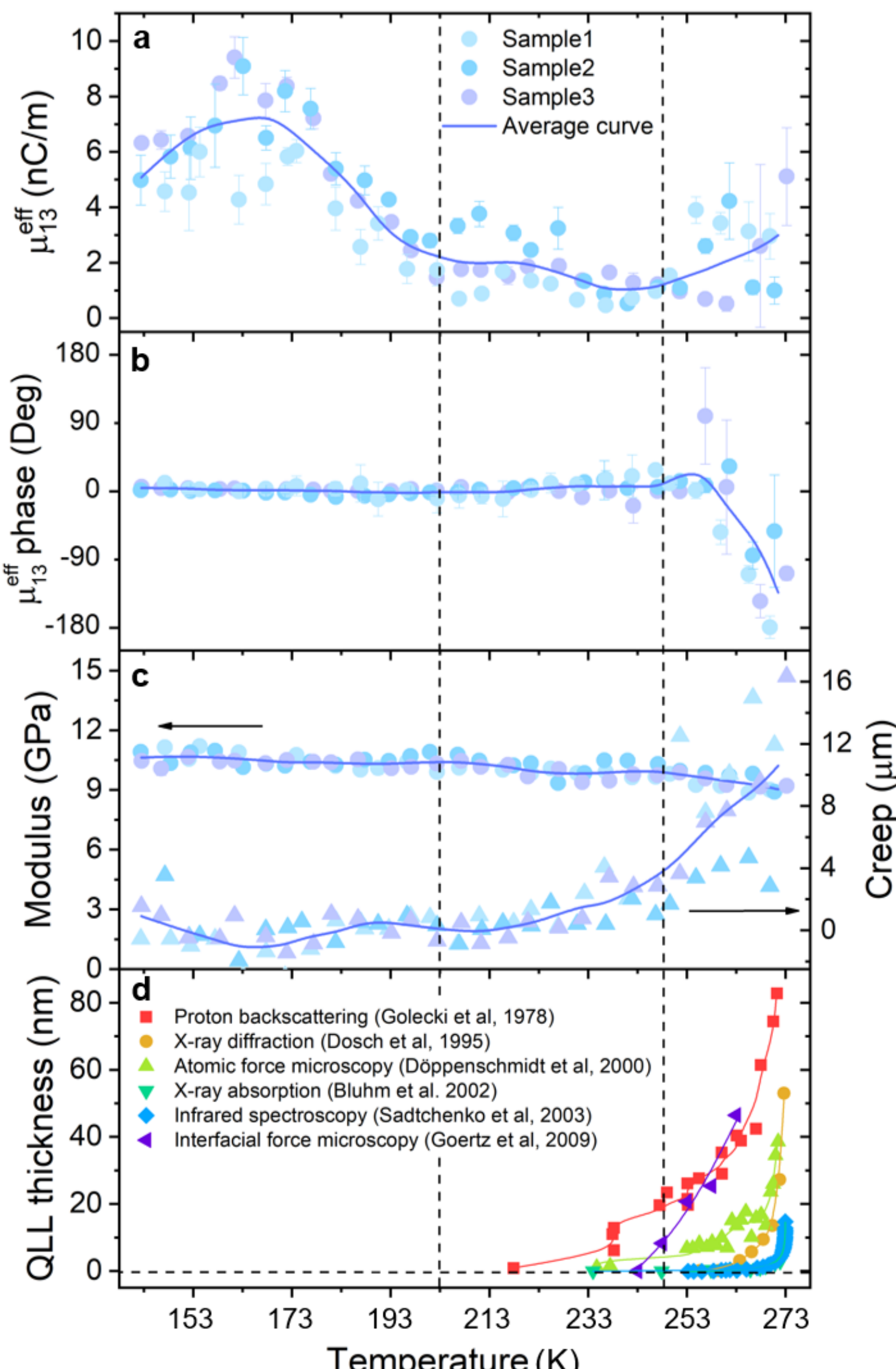

**Fig. S4**. **a**, The flexoelectric coefficient with Au electrodes versus temperature. **b**, The phase angle between displacement and polarization charge with Au electrodes versus temperature. **c**, The modulus and the creep displacement in the first ten seconds of loading versus temperature. **d**, The reported QLL thickness versus temperature measured with different techniques: X-ray absorption[24], Atomic force microscopy[25], Proton backscattering[26], X-ray diffraction[27], Infrared spectroscopy[28], Interfacial force microscopy[29].

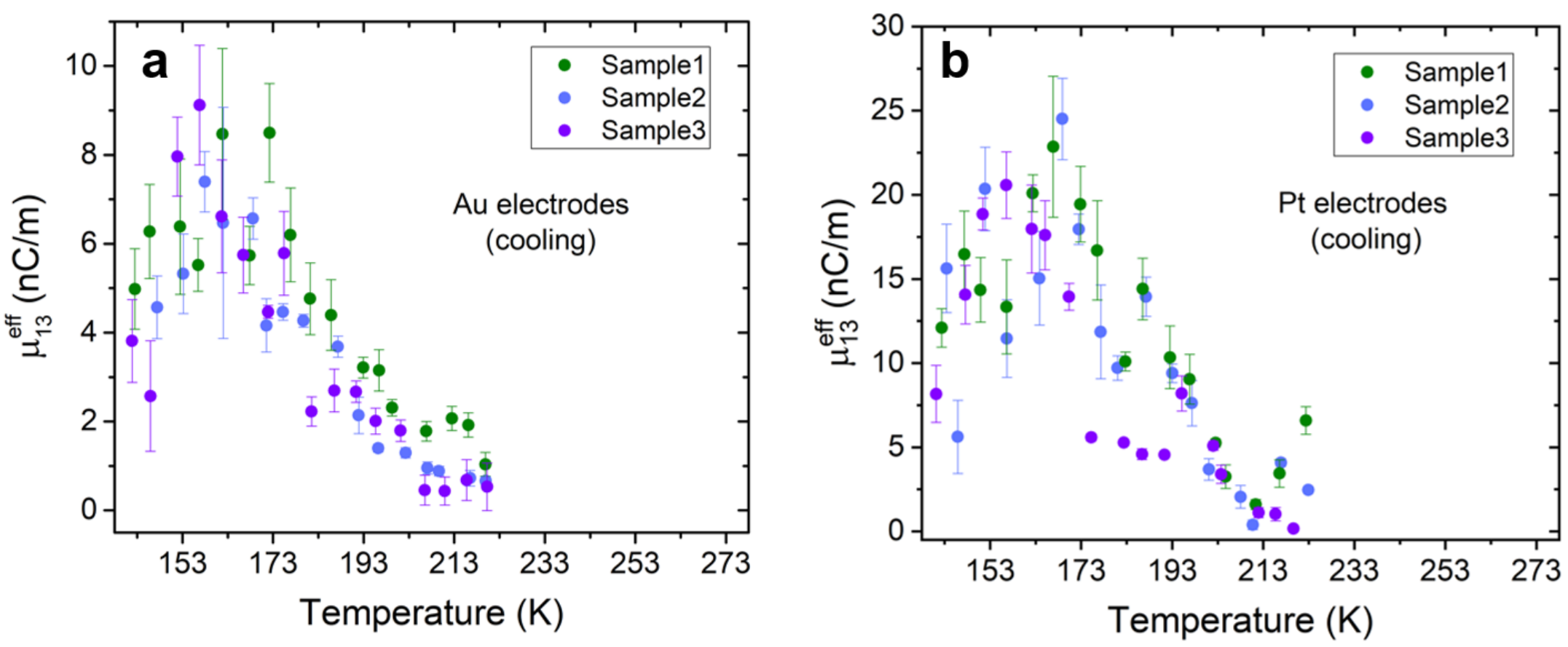

**Fig. S5.** The flexoelectric coefficient measured on cooling for samples with (a) Au electrodes and (b) Pt electrodes. The error bars represent the standard error from linear regressions.

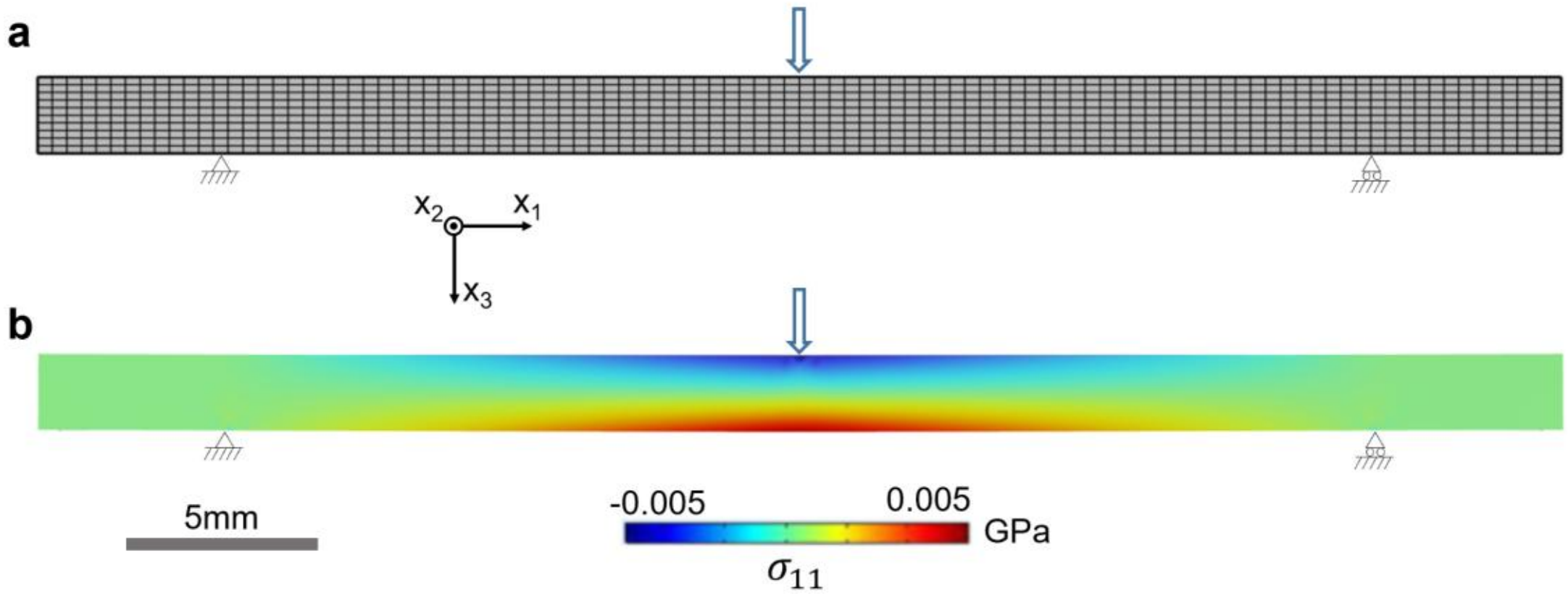

**Fig. S6.** Finite-element simulation of stress distribution in our sample under the maximum force (-2.5N).

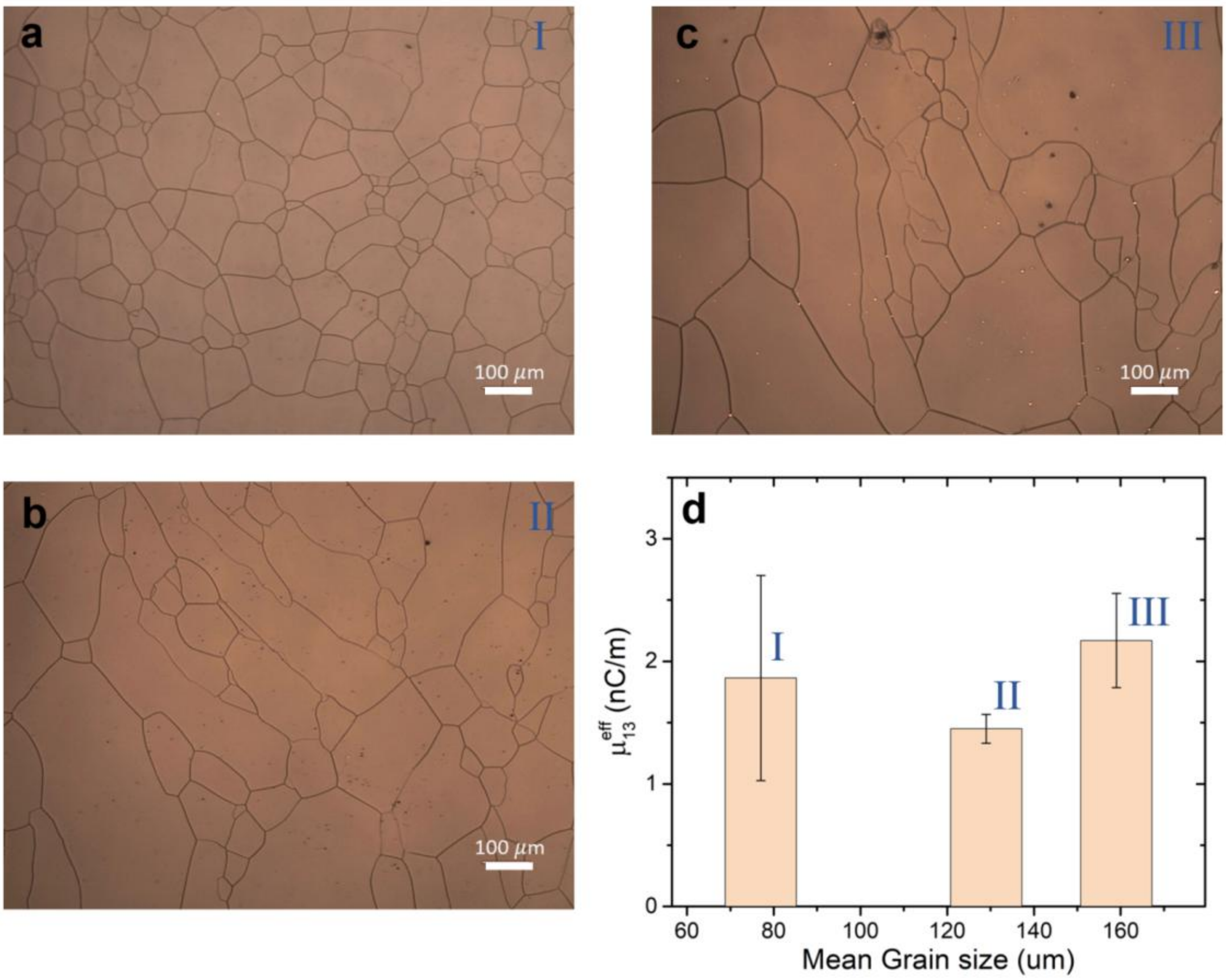

**Fig. S7**. Optical image of the grains on the surface of our ice sample (a) before annealing, (b) after annealing for half hour at 267 K and (c) after annealing for one hour at 267 K. d, The grain-size dependence of the flexoelectric coefficient measured at 233K.

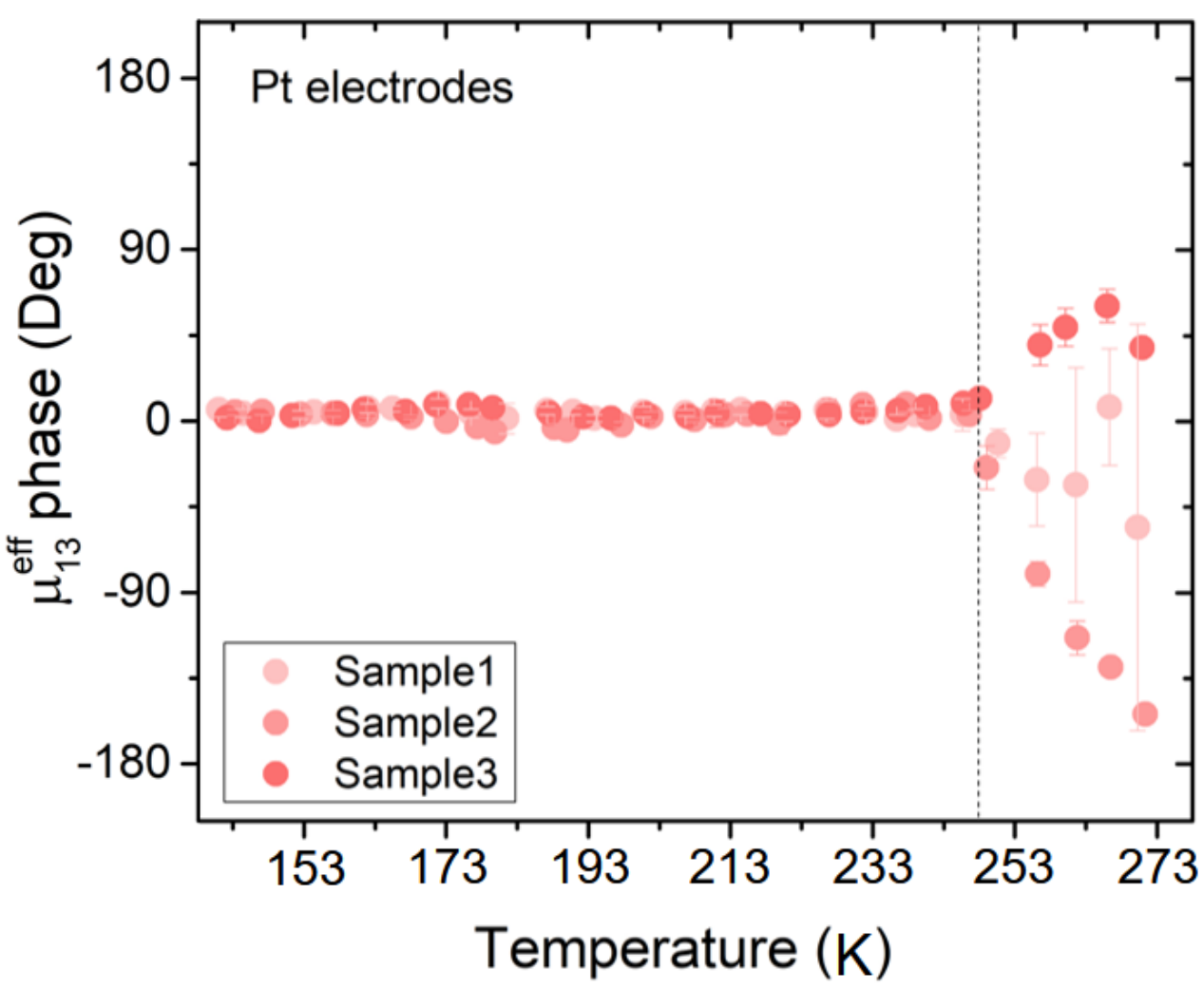

Fig. S8. The phase angle between displacement and charge measured in three ice samples with Pt electrodes on heating. The error bars represent the standard deviation from the average of multiple measurements.

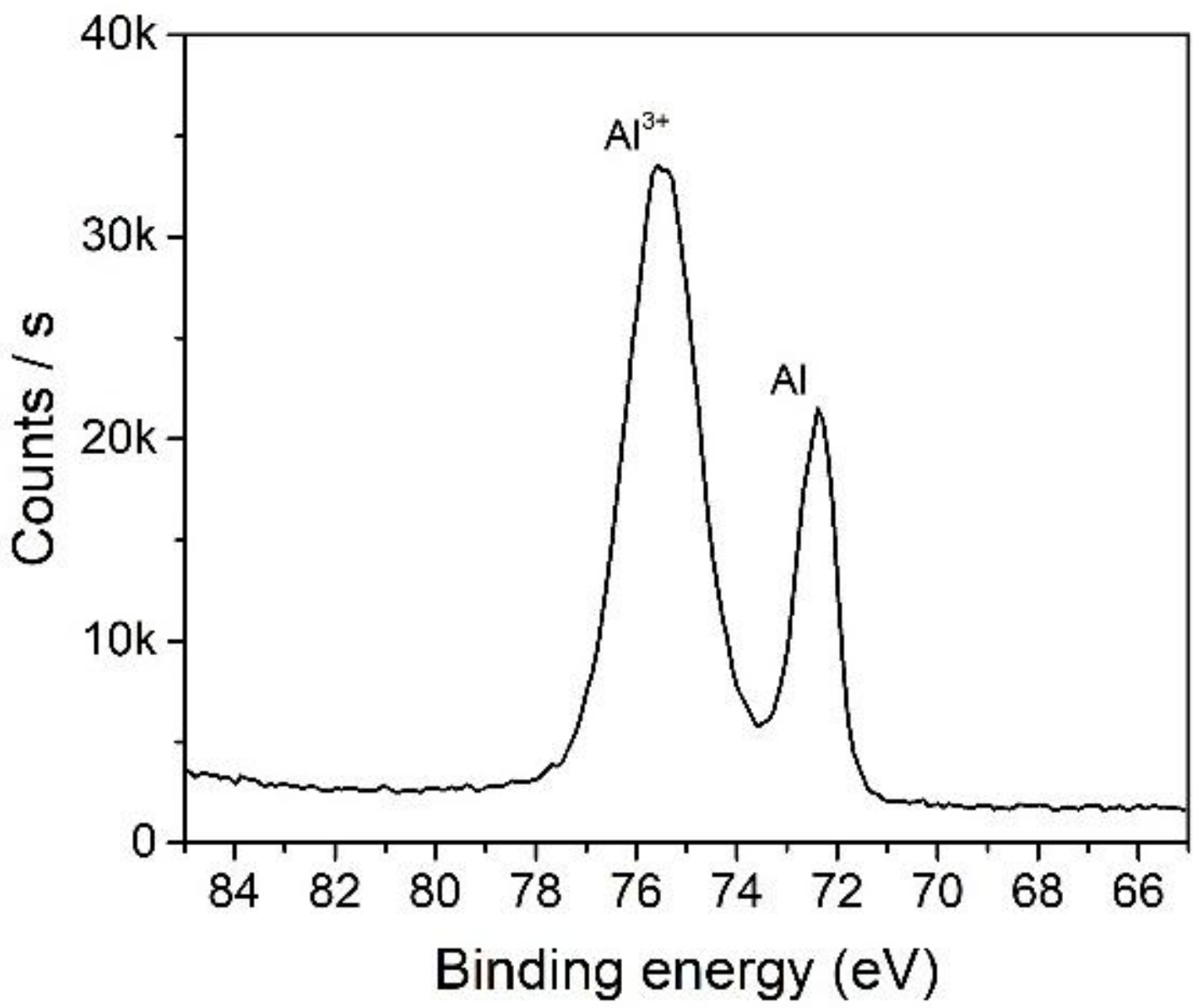

Fig. S9. XPS result showing the presence of oxidized $Al^{3+}$ (attributed to $Al_2O_3$) at the surface of the Al electrode

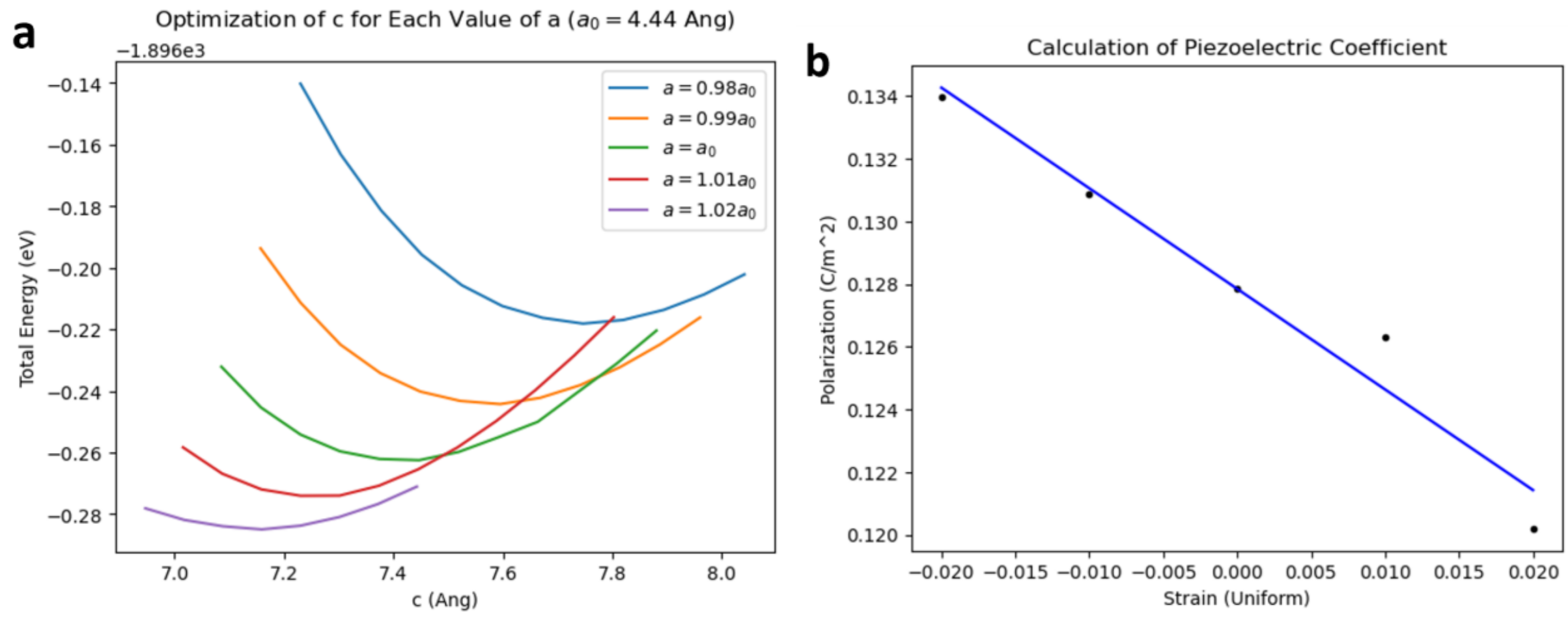

Fig. S10. Calculation of the transverse piezoelectric constant of ice XI.

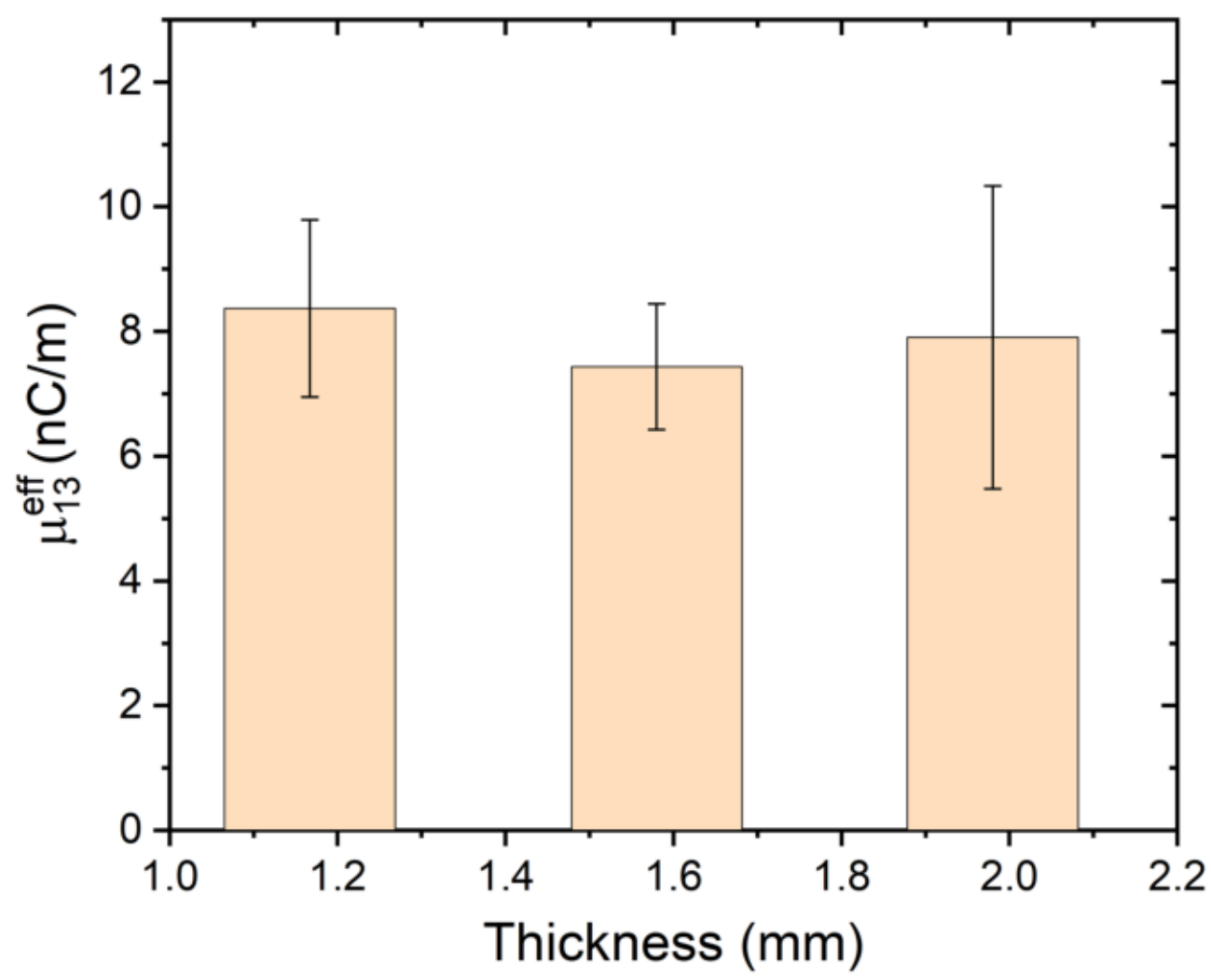

Fig. S11. Thickness dependence of $\mu_{13}^{eff}$ for ice with Au electrodes measured at ~160K.

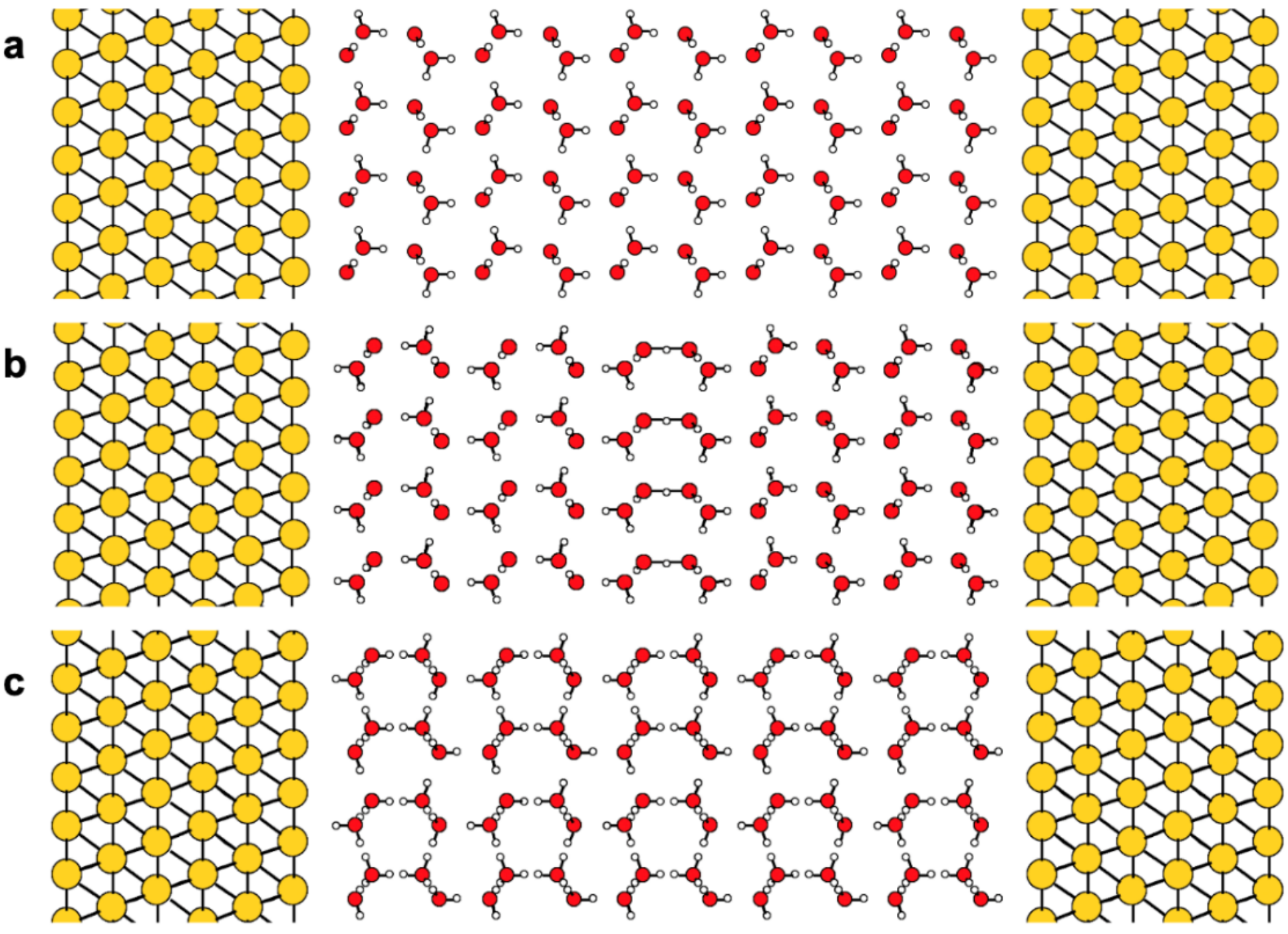

**Fig. S12.** Illustration of ice XI-Au interface with **(a)** one O-interface and one H-interface, and **(b)** two H-interfaces with a layer of defects in the middle. (c) Illustration of ice Ih-Au interface.

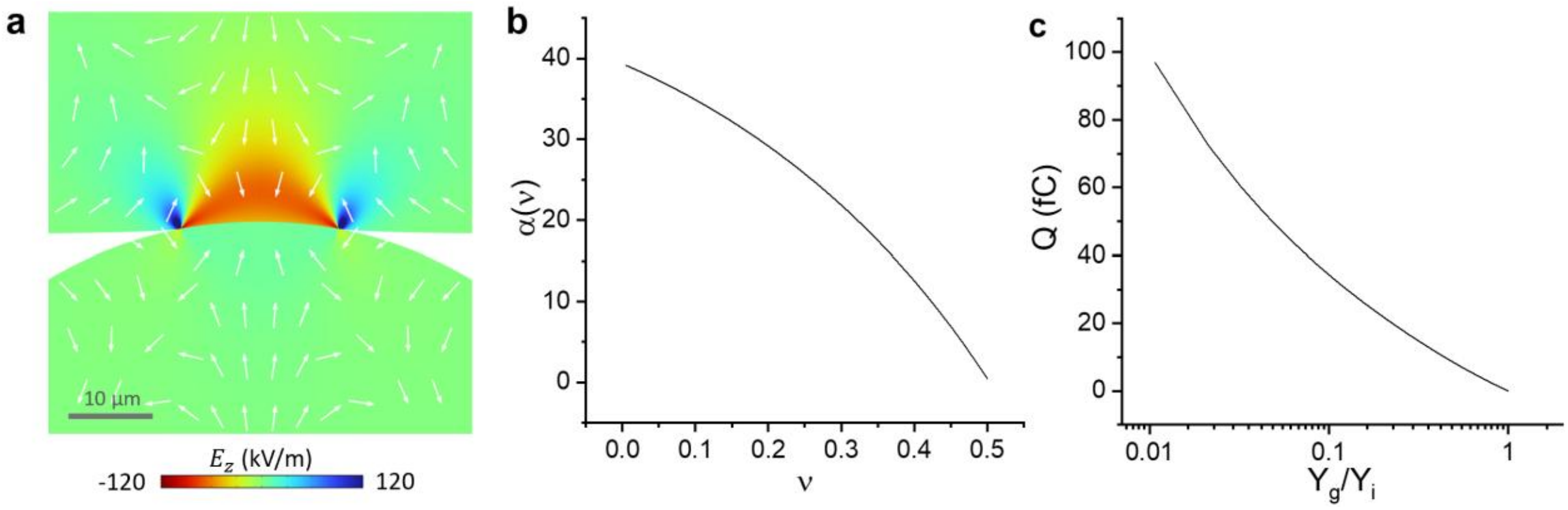

**Fig. S13. (a)** The calculated depolarization field near the contact interface between the graupel and ice particle**. (b)** The relationship between Poisson's ratio $\nu$ and $\alpha(\nu)$, a function defined in Equation (S26). **(c)** $Q$ versus the ratio between graupel's and ice's Young's modulus. See Table S1 for adopted parameters.

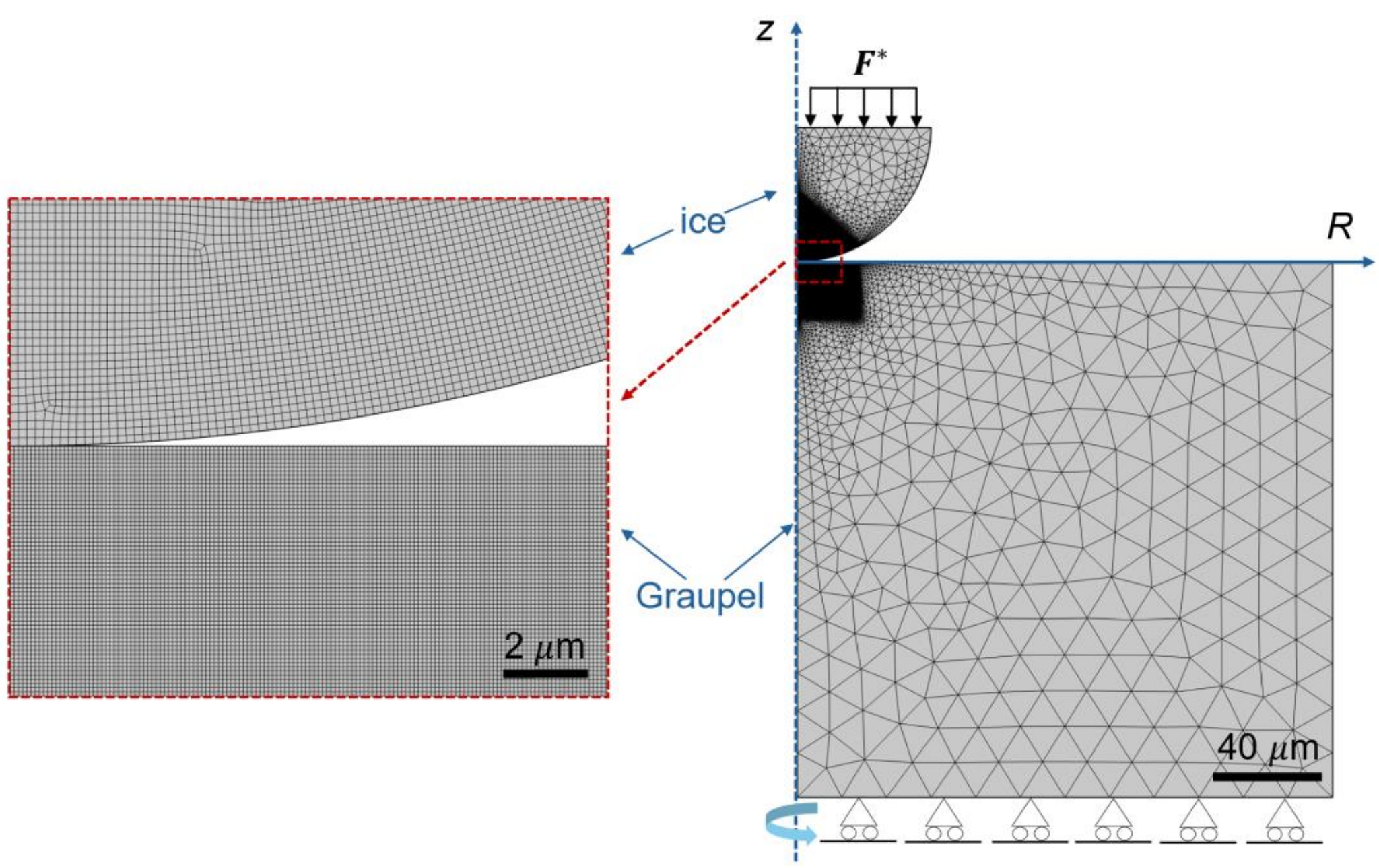

**Fig. S14.** Finite element model

| $\boldsymbol{Y_i}$ (GPa) | $\boldsymbol{Y_g}$ (GPa) | $\boldsymbol{\nu=\nu_i=\nu_g}$ | $\rho$ **(kg/m³)** |
|---|---|---|---|
| 9.33[43] | 1[43,53] | 0.325[43] | 916.7[43] |
| $\boldsymbol{\xi_r}$ | $\boldsymbol{\mu_{13}^{eff}}$ (nC/m) | $\boldsymbol{R_i}$ (μm) | $\boldsymbol{v_r}$ (m/s) |
| 100[43] | 1.14±0.13 | 50 | 0~22 |

**Table S1.** Parameters used in theoretical calculations. The velocity $v_r$ is varied within 0~22 m/s for Fig. 5c, and is fixed at 8 m/s—the condition used in Gaskell's experiments (see Section S10)—for Fig. 5b and Fig. S13a~c. In Fig. S13b and S13c, the parameters $\nu$ and $Y_g$ are varied respectively while all other parameters remain constant described here.